%% file: main.tex
\documentclass[final,5p,times,twocolumn,authoryear]{elsarticle}

\journal{Computers $\&$ Security Journal}

\makeatletter
\def\ps@pprintTitle{%
  \let\@oddhead\@empty
  \let\@evenhead\@empty
  \def\@oddfoot{\footnotesize\itshape
         {Accepted Preprint in Computers and Security Journal} \hfill\today}%
  \let\@evenfoot\@oddfoot
}
\makeatother

\usepackage{amssymb}
\usepackage{lipsum}

\usepackage{amsmath,amsfonts}
\usepackage{algorithmic}
\usepackage[ruled,vlined]{algorithm2e}
\usepackage{textcomp}
\usepackage{xcolor}
\usepackage{array}
\usepackage{tabularx}
\usepackage{booktabs}
\usepackage{float}
\usepackage{enumerate} 
\usepackage{multirow}
\usepackage[caption = false]{subfig}
\usepackage{fancyvrb}

\usepackage{xcolor}
\usepackage[colorlinks]{hyperref}

\usepackage{paralist}

\usepackage{epsfig}
\usepackage{booktabs}
\usepackage[toc]{appendix}

\newcommand{\kms}{km\,s$^{-1}$}

\DeclareMathOperator*{\argmax}{arg\,max}  

\newcommand{\toolname}{\textsc{MAlign}}
\newcommand{\nucleicseq}{nucleotide sequence}
\newcommand{\consensusseq}{consensus sequence}

\usepackage[T1]{fontenc}
\usepackage[breakable, theorems, skins]{tcolorbox}
\tcbset{enhanced}

\begin{document}

\begin{frontmatter}



\title{\toolname : Explainable Static Raw-byte Based Malware Family Classification \\ using Sequence Alignment}


\affiliation[first]{organization={University of Maryland},
            city={College Park},
            state={Maryland},
            country={United States}}
            
\affiliation[second]{organization={Bangladesh University of Engineering and Technology (BUET)},
            city={Dhaka},
            country={Bangladesh}}

\affiliation[third]{organization={ICSI, University of California, Berkeley},
            country={United States}}
            
\affiliation[fourth]{organization={Avast},
            country={United States}}

\author[first,second]{Shoumik Saha\corref{Cor1}}\ead{smksaha@umd.edu}
\author[third,fourth]{Sadia Afroz}
\author[second]{Atif Hasan Rahman\corref{Cor1}}\ead{atif@cse.buet.ac.bd}
\cortext[Cor1]{Corresponding authors}

\begin{abstract}
\input{files/abstract}
\end{abstract}




\end{frontmatter}




\section{Introduction}

\input{files/intro}

\section{Related Work}
\input{files/related_work}

\section{Background}
\input{files/background}

\section{Methods}
\input{files/our_method}

\section{Evaluation}
\input{files/results}

\section{Explainability}
\input{files/explainability}

\section{Robustness}
\input{files/robustness}

\input{files/app_runtime}

\section{Discussion}
\input{files/discussion}

\section{Conclusion}
\input{files/conclusion}

\section*{Code Availability}
Our code can be found here - \url{https://github.com/ShoumikSaha/Malign}

\section*{CRediT authorship contribution statement}
\begin{itemize}
    \item \textbf{Shoumik Saha: } Conceptualization, Data curation, Formal Analysis, Methodology, Software, Validation, Visualization, Writing – original draft, Writing – review \& editing
    \item \textbf{Sadia Afroz: } Resources, Supervision, Writing – review \& editing
    \item \textbf{Atif Hasan Rahman: } Conceptualization, Formal Analysis, Investigation, Resources, Supervision, Writing – review \& editing
\end{itemize}

\section*{Acknowledgements}
We would like to express our sincere gratitude to Erin Avllazagaj for his review and feedback. He helped us to improve the paper by analyzing some signatures.

\bibliographystyle{elsarticle-harv} 
\bibliography{main}
\newpage
\onecolumn
\appendix

\input{files/app_models}
\newpage
\input{files/app_results}

\newpage
\input{files/app_explainability}

\input{files/app_robustness}

\newpage
\input{files/app_seq_analysis}







\end{document}

%% file: files/abstract.tex
For a long time, malware classification and analysis have been an arms-race between antivirus systems and malware authors. Though static analysis is vulnerable to evasion techniques, it is still popular as the first line of defense in antivirus systems. But most of the static analyzers failed to gain the trust of practitioners due to their black-box nature.
We propose \toolname, a novel static malware family classification approach inspired by genome sequence alignment that can not only classify malware families but can also provide explanations for its decision. \toolname\ encodes raw bytes using nucleotides and adopts genome sequence alignment approaches to create a signature of a malware family based on the conserved code segments in that family, without any human labor or expertise.
We evaluate \toolname\ on two malware datasets, and it outperforms other state-of-the-art machine learning-based malware classifiers (by $4.49\% {\sim} 0.07\%$), especially on small datasets (by $19.48\% {\sim} 1.2\%$). Furthermore, we explain the generated signatures by \toolname\ on different malware families illustrating the kinds of insights it can provide to analysts, and show its efficacy as an analysis tool. 
Additionally, we evaluate its theoretical and empirical robustness against some common attacks. In this paper, we approach static malware analysis from a unique perspective, aiming to strike a delicate balance among performance, interpretability, and robustness.

%% file: files/intro.tex
\label{intro}

Over the past decade, malware has been growing and spreading exponentially. For example, the AV-Test institute registers more than 450K new malware and potentially unwanted programs (PUP) every day \citep{malware_stat}. They found 16 million new malware in just the first 3 months of 2023.\footnote{{https://portal.av-atlas.org/malware}}

To detect this rising number of malware at scale, static analysis is used as the first line of defense by all commercial antivirus (AV) systems, because it requires less computational power and time than dynamic analysis. However, dynamic analysis allows security analysts to perform more in-depth analysis by executing the malware in a sandbox or cloud, and capturing its behavior and activity \citep{or2019dynamic, anderson2011graph, avllazagaj2021malware}. Additionally, static analysis is comparatively easier to evade than dynamic ones.

Like all other fields, machine learning (ML) is getting adapted to security too, including malware detection and family classification~\citep{schultz2000data, nataraj2011malware, pascanu2015malware, kalash2018malware, shahzad2011accurate, lu2019malware, malnet, ahmadi2016novel, malconv}. 
ML has proved helpful in detecting the rising number of malware at scale. In fact, most commercial malware detectors use machine learning for their static analysis nowadays.
However, these deep-learning based black-box models lack \textit{interpretability}, and conventional explanation methods for machine learning~\citep{zhang2019should} are not suitable in security~\citep{guo2018lemna}. Consequently, security practitioners cannot fully trust these models, especially in safety-critical applications where the ability to explain a classification error is crucial.

Unfortunately, while a large body of work has gone into designing highly accurate static malware classifiers by outperforming each other over the past few years, very little work has been done considering the \textit{explainability} aspect. Existing works~\citep{arp2014drebin, kinkead2021towards, melis2018explaining, liu2022explainable, kumar2018effective} on explainable malware classifier are done on extracted features, e.g., requested and used permissions, API calls, network addresses, etc., rather than analyzing the executable code directly. Thus, they can only provide explainability at a higher-level semantics, while the potential insights and new knowledge that could be extracted from low-level input features remain unexplored. This untapped information could include valuable findings such as common practices among malware authors, prevalent obfuscation techniques, and other critical insights essential for analysts.


Another shortcoming of current static malware classifiers is that they are \textit{vulnerable} to different adversarial attacks ~\citep{song2020mab, suciu2019exploring, demetrio2021functionality, ceschin2019shallow, kolosnjaji2018adversarial, pe11,pe13}. 
Though these adversarial attacks use different techniques and work in different settings, fundamentally, they fall under one umbrella -- adding or modifying bytes (from benign files or generated using gradient approaches) without altering the malware semantics. Consequently, there have been works on proposing defenses for such attacks, e.g., monotonic classifier ~\citep{fleshman2018non}, adversarial training ~\citep{lucasadversarial}, etc,. Unfortunately, most of these defenses suffer from low standard accuracy and can provide robustness to a limited number of attacks; which introduces a trade-off between robustness and accuracy.

Moreover, end-to-end deep-learning based models are \textit{data hungry}~\citep{marcus2018deep}. The more data they get, the better they learn the features. However, this data hunger poses a concerning challenge in the context of malware family classification, particularly in the detection of a new variant. When a new malware variant or family emerges, it takes time for AV vendors to collect enough samples from the wild. Unfortunately, the prevalence of such new variants is on a steady rise. For example, in 2020, there was a 62\% rise in detected malware variants~\footnote{https://dataprot.net/statistics/malware-statistics/}. In March 2022, there was a new record of discovering 60,000 new malware variants~\footnote{https://www.comparitech.com/antivirus/malware-statistics-facts/}. As a result, these deep-learning based approaches fail to perform at their fullest for such new malware variants due to their insufficient training set size.

All of these issues raise a crucial research question:

\textit{"Can we design a novel static analysis technique that can effectively mitigate the shortcomings (non-interpretability, vulnerability, data-hungry) mentioned above?"}


In this study, we focus on a raw-byte-based static malware family classification technique as a potential answer to this question. By exploring this approach, we aim to enhance the explainability and robustness of static analysis in combating the ever-evolving landscape of malware threats.

\textbf{Our Design.}
We propose \toolname, a novel static malware family detection method that operates directly on the raw bytes from executables. Unlike conventional static analyzers, our approach leverages the sequence alignment mechanism from bioinformatics and incorporates it with a machine learning model. The rationale behind adopting such an alignment mechanism is the indistinct similarity between genome sequences and malware executables. Genomes contain critical regions for the survival of the organism, such as protein-coding genes where mutations (or changes in genes) can be lethal. Similarly, malware families contain malicious code blocks associated with the malicious functionalities of that family, and altering them can change their semantics or corrupt the files. Embracing the analogy between these two distinct domains, we capitalize on the advanced alignment methods developed in bioinformatics. Our rationale is that -- this alignment technique will capture code blocks that are common in one malware family the way it does for genomes in species. 

The alignment technique identifies the common blocks, i.e., consensus sequences, per family, and estimates the degree of conservation at each location by processing the generated alignments. Then we generate a conservation score for each of these blocks depending on their conservation and frequency in that family. In this context, these aligned blocks along with their scores can be considered as the features for that corresponding malware family. However, there can be aligned common blocks that are shared in all malware families (for example, common PE header), and hence, they will not represent any family-specific functionalities. 
Therefore, we train a classifier on these generated conservation scores from these aligned blocks, including different families, so that the classifier learns the importance of these generated blocks for a specific family. 
During inference, to classify whether a new malware belongs to a family, we generate the alignment of this new malware with the previously generated common blocks of that family, and use it to compute a set of alignment scores. These scores are then fed to the classifier model to classify the new malware to its family.

Additionally, our adopted alignment approach can effectively handle addition, subtraction, substitution, and re-ordering in sequences, which provides \textit{better robustness} to perturbation than conventional techniques. 
Furthermore, the use of such a backtrackable alignment approach with an \textit{interpretable} ML model enables us to map its classification decision to the responsible suspicious code blocks. As a result, \toolname\ is trustworthy for its decision which is highly crucial for malware applications.

\textbf{Evaluation.}
We evaluate \toolname\ for detecting malware families on two datasets: (i) Kaggle Microsoft Malware Classification Challenge (Big 2015)~\citep{microsoftdataset2015}, and (ii) Microsoft  Machine  Learning  Security  Evasion  Competition (2020)~\citep{mlsec2020}. In comparison to the state-of-the-art methods such as MalConv~\citep{malconv}, CNN-based model~\citep{kalash2018malware}, and Feature-Fusion model~\citep{ahmadi2016novel}, our approach has higher accuracy, and the difference in performance is more noticeable when datasets are small. Using our auto-generated aligned blocks, i.e. consensus sequences, for each family, we backtracked to the exact code blocks that were responsible for our model decision. Thus, we were able to discover suspicious functionalities from malware samples, making our model explainable. Moreover, we theoretically prove that \toolname\ is robust by-design to gradient attacks. We also evaluate the empirical robustness by generating adversarial malware using a gradient-based patch-append attack.

In summary, our main contributions are:~
\begin{enumerate}

    \item \textbf{Sequence alignment based approach.} We propose \toolname\ by adopting sequence alignment concepts from bioinformatics into the static malware family classifier. We used a multiple sequence alignment-based tool to find conserved regions in raw binary executables
    (Section \ref{sec:method}).
    
    \item \textbf{High accuracy.} We evaluated \toolname\ on two datasets, and it outperforms other state-of-the-art static detection methods. Besides, the performance gap gets higher with a lower amount of training data (Section \ref{sec:evaluation}). 
    
    \item \textbf{Explainable by-design.} Due to our design choice, \toolname\ is backtrackable and explainable for its decision.
    We explore this in-depth and identify responsible code-blocks for multiple malware families. 
    We share some of the interesting findings and insights that can help the security community
    (Section \ref{sec:explainability}). 
    
    \item \textbf{Robust by-design.} We mathematically prove that our design choice makes \toolname\ robust against gradient attacks. We also evaluate its empirical robustness against gradient-based patch attacks (Section \ref{sec:robustness}).
\end{enumerate}

%% file: files/related_work.tex
\label{related_work}

\textbf{Machine Learning based classifiers. }
To counter the increasing amount of malware and detect them, several methods and techniques have been developed over the years. In the early days, ~\cite{wressnegger2017automatically} and ~\cite{zakeri2015static} proposed a signature-based approach using static analysis. Later, a dynamic approach - malware detection by analyzing the malware behavior,  was proposed by ~\cite{martignoni2008layered} and ~\cite{willems2007toward}. In recent times, machine learning based techniques are also being used to classify malware.
~\cite{schultz2000data} first proposed a data mining technique for malware detection using three different types of static features. Subsequently, ~\cite{nataraj2011malware} proposed a malware classification approach based on image processing techniques by converting the bytes files to image files. Later,~\cite{kalash2018malware} improved on ~\citep{nataraj2011malware} by developing M-CNN using convolutional neural networks (CNN). Besides CNN, RNN has also been used for malware analysis.~\cite{shahzad2011accurate} and~\cite{lu2019malware} proposed techniques with LSTM using opcode sequences of malware. \cite{santos2013opem} proposed a hybrid technique by integrating both static and dynamic analysis. Subsequently, ~\cite{malnet} developed MalNet using an ensemble of CNN, LSTM and by extracting metadata features, while Ahmadi {et al.}~\cite{ahmadi2016novel} extracted and selected features of malware depending on the importance and applied feature fusion on them.
Recently, ~\cite{malconv} developed a state-of-the-art technique MalConv using only the raw byte sequence as the input to a CNN-based deep-learning model. 

\textbf{Sequence Alignment based Classifiers. }
In the past, sequence alignment based approaches have been used for malware analysis by a number of researchers~\citep{chen2012multiple,narayanan2012effects, naidu2014further, kirat2015malgene, naidu2016needleman, cho2016malware, kim2019improvement}.
\cite{chen2012multiple} used multiple sequence alignment to align 
computer viral and worm codes of variable lengths to identify invariant regions. This approach was subsequently enhanced later~\citep{narayanan2012effects, naidu2014further,naidu2016needleman}.
However, these approaches were not scalable enough to train on large malware sets and did not evaluate the robustness of their models, unlike our work.

Sequence alignment has also been applied on system call sequences of malware to extract evasion signatures and to cluster samples by~\cite{kirat2015malgene}, classification of malware families by~\cite{cho2016malware}, and for malware detection, classification by~\cite{DANGELO2021107234}, and visualization by~\cite{kim2019improvement}. 
Along with this line of work, there are existing works that proposed ensemble detection leveraging sequence alignment on API sequences~\citep{ficco2021malware}. 
All these approaches require access to API call sequences which require dynamic analysis and execution in a sandbox, virtual machine, or host. However, our proposed method is based on static analysis and only takes the raw bytes of a file which does not require any execution in a secondary machine.

~\cite{drew2016polymorphic} utilized another approach developed by the bioinformatics and
computational biology community and used gene or sequence classification methods for malware classification. The method is based on extracting short words i.e. $k$-mers from sequences and calculating the similarity between sequences based on the set of words present in them. Although the method is efficient, it does not fully utilize the information provided by stretches of conserved regions in malware and is not suitable for identifying critical code blocks in malware. In contrast, our method takes the whole raw-byte executable into consideration and so, can explain its decision by backtracking to the responsible code blocks.

\textbf{Explainability of Models. }
Though there has been a large body of work on classifying and detecting malware, not much work has been done on explaining and interpreting the model decision. However, there are a few works that explored explainability, e.g., DREBIN ~\citep{arp2014drebin}, ~\cite{melis2018explaining, melis2022gradient}, ~\cite{backes2017luna}, ~\cite{kumar2018effective},  ~\cite{kinkead2021towards}, MalDAE ~\citep{han2019maldae}, XMAL ~\citep{wu2021android}, etc. Unfortunately, all of these works were on Android malware using higher-level features like API calls, requested permissions, opcode sequences, etc., from manifest and disassembled code. In contrast, \toolname\ takes raw-bytes from executables (Windows) and can explain the decision from that. 

On the other hand, recent works in the machine learning community like LIME ~\citep{zhang2019should}, SHAP ~\citep{lundberg2017unified} propose strategies to interpret a decision of a blackbox model. However, these strategies assume that the local area of the classification boundary is linear. Later ~\cite{guo2018lemna} proposed LEMNA showing that -- when the decision boundary is non-linear, previous approaches would produce errors which makes them unsuitable for security applications. So, they solved this by using a mixture regression model enhanced by fused lasso. Though such models have played a great role in the explainability domain and inspired us, they have to be implemented and trained separately on a classifier model, which requires human engineering (like hyperparameter tuning) and adds computational overhead. However, our proposed method, \toolname\ has a pipeline that is interpretable from top to bottom, i.e., easy to backtrack from decision output to problem space. Thus, it does not require any other model on top of it to explain its decision. Moreover, we generate signatures for each family that are easily interpretable by practitioners (see Section \ref{sec:explainability}).  

\textbf{Adversarial Robustness. }
Continuous circumvention from malware authors and defense against such evasion by AV systems have turned into a never-ending cat-and-mouse game. For example, in the whitebox setting, ~\cite{suciu2019exploring},  ~\cite{kolosnjaji2018adversarial}, ~\cite{kreuk2018deceiving}, ~\cite{grosse2016adversarial}, ~\cite{pe15}, and in the blackbox setting,  ~\cite{demetrio2021functionality}, ~\cite{ceschin2019shallow},  ~\cite{fleshman2018static}, ~\cite{pe18}, ~\cite{pe17}, ~\cite{pe19}, showed that adversarial attacks are effective against machine learning (ML) models in multiple ways. Also, ~\cite{pe14}, ~\cite{song2020mab} proposed a reinforcement based approach to rewrite a binary for evading ML models.

Prior work proposed two main ways to improve the adversarial robustness of malware detectors: adversarial training and robustness by design. 
Adversarial training, where a malware detector is trained with adversarial examples, is one of the most commonly used approaches to improve adversarial robustness~\citep{bai2021recent, grosse2017adversarial,zhang2021enhanced,al2018adversarial}. However, such approaches hamper the classification accuracy, add computational overhead, and rely on the availability of enough adversarial samples, which may not always be the case in the fast-changing malware world \citep{lucasadversarial}.
So, in this work, we focus on the latter one -- robustness by design.

Robustness by design approaches build classifiers to eliminate a certain class of adversarial attacks. 
For example, \cite{chen2020training} proposed learning PDF malware detectors with verifiable robustness properties, \cite{saha2023drsm} proposed a detector adopting the de-randomized smoothing scheme with certified robustness, \cite{incer2018adversarially} trained an XGBoost based malware detector with the monotonicity property, \cite{fleshman2018non} trained a non-negative network to ensure that an adversary cannot decrease the classification score by adding extra content, etc. However, these approaches had to compromise the detection accuracy to some extent. For example, the non-negative MalConv model's detection accuracy drops from 94.1\% to 89.4\%, while \toolname\ can hold its performance as well as its robustness.

Moreover, recent works~\citep{lucas2021malware, lucasadversarial} demonstrate the binary-diversification technique against new and old ML defenses. It is fundamental that -- no ML model is fully robust against all attacks, and so it is better to not totally rely on ML models. So, we propose \toolname\ that incorporates sequence alignment in front of ML models which makes it non-invertible and non-differentiable. We evaluate our model against a gradient-based attack and theoretically prove that it is robust-by-design.

%% file: files/background.tex
Sequence alignment is a widely studied problem in bioinformatics to find similarities among DNA, RNA, or protein sequences, and to study evolutionary relationships among diverse species. It is the process of arranging sequences in such a way that regions of similarity are \emph{aligned}, with gaps (denoted by `-') inserted to represent insertions and deletions in sequences. For example, an alignment of the sequences \texttt{ATTGACCTGA} and \texttt{ATCGTGTA} is shown below where the regions denoted in black, characterized by identical characters, are matched, whereas the red and blue regions denote substitutions i.e. point mutations, and insertions or deletions during the evolutionary process, respectively.

\begin{center}
\begin{BVerbatim}[commandchars=\\\{\}]

AT\textcolor{blue}{TGA}C\textcolor{red}{C}TG\textcolor{blue}{-}A
AT\textcolor{blue}{---}C\textcolor{red}{G}TG\textcolor{blue}{T}A
\end{BVerbatim}
\end{center}

In sequence alignment, matches, mismatches, and insertions-deletions (in-dels) are assigned scores based on their frequencies during evolution and the goal is to find an alignment with the maximum score.
The problem of finding an optimal alignment of the entire sequences (global alignment) and that of finding an optimal alignment of their sub-sequences (local alignment) can be solved by dynamic programming using the Needleman-Wunsch~\citep{needleman-wunsch} and the Smith-Waterman~\citep{smith-waterman} algorithms, respectively. While the algorithms can be used to align more than two sequences, the running time is exponential in the number of sequences. To address the tractability issue, a number of tools have been developed~\citep{clustal, mafft, muscle}, that use heuristics to solve the multiple sequence alignment (MSA) problem~\citep{msa}.

However, in addition to point mutations and short insertions-deletions, large scale genome rearrangement events take place during evolution. Such genome rearrangement events include reversal of a genomic segment (inversion), the shuffling of the order of genomic segments (transposition or translocation), duplication and deletion of segments, etc. Although the aforementioned tools are unable to deal with genome rearrangements, methods such as MUMmer~\citep{mummer} can perform alignment of two sequences in the presence of rearrangements, whereas Mauve~\citep{mauve}, Cactus~\citep{cactus}, etc.\ can handle multiple sequences.

Recently, ~\cite{progressivecactus} and ~\cite{sibeliaz} have developed Progressive Cactus and SibeliaZ respectively, that can align hundreds to thousands of whole-genome sequences that may include rearrangements.
The tools identify similar sub-sequences in the sequences from different species to create blocks of rearrangement-free sequences, and then perform a multiple sequence alignment of the sequences in each block.   

Since adversaries can modify malware relatively easily by changing orders of blocks of codes without changing malware semantics, it is important that the tool used to align malware sequences is robust to such rearrangements in code. Here we use SibeliaZ to align malware sequences to identify conserved blocks of codes and calculate a conservation score of the blocks for malware family detection and classification (see Section \ref{sec:method}). It is worth noting that the blocks of codes identified need not be fully conserved, i.e. there can be modifications, insertions, and deletions of a small number of instructions within the blocks, making it robust to adversarial attacks to some extent (see Section \ref{sec:robustness}).

%% file: files/our_method.tex
\label{sec:method}

\begin{figure*}[h]
    \centering
    \includegraphics[width=1.0\textwidth]{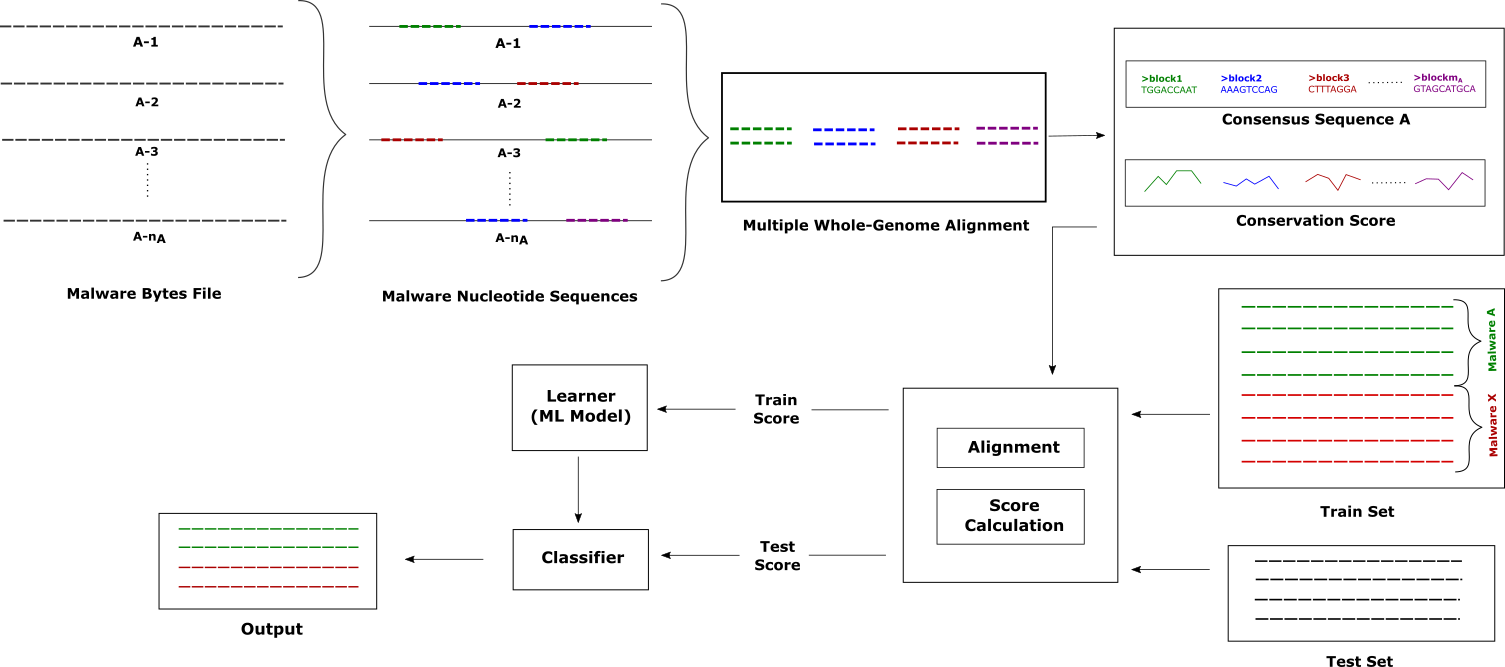}
    \caption{Overview of \toolname.  (1) The malware bytes files (executables) from malware of a particular family are first converted to nucleotide sequence files. (2) Then the malware nucleotide sequences are aligned using a multiple sequence alignment tool SibeliaZ. It first identifies similar sequences in different malware to form blocks. Highly similar sequences (colored sequences) can be in different orders in different files. The sequences in each block are then aligned.    (3) The aligned sequences in each block are used to construct consensus sequences and conservation scores are calculated for each conserved block. (4) Then two sets of sequences - one corresponding to the malware family of interest and the other corresponding to non-malware or malware from other families are aligned to the consensus sequences and the degrees of conservation of each conserved block in the training sequences are estimated. (5) Finally a machine learning model is learnt to classify sequences based on the alignment scores of the sequences with the blocks. To classify new instances, sequences are aligned to the consensus sequences of the blocks and alignments are scored. The scores are then used as features for the class prediction.}
    \label{fig:overview}
\end{figure*}

\subsection{\textbf{Overview}}

To classify known malware family and their variants, here we propose a malware family classification system based on multiple whole-genome alignment. The basic building block of the method is a binary classification system that can predict whether an executable belongs to a particular malware family or not. The input to this binary classifier is a training set consisting of positive samples i.e. malware from a particular family, and negative samples, which can be benign or malware from other families. 

\begin{algorithm}[ht]
\footnotesize
\DontPrintSemicolon
  
  \KwIn{Training set, $\mathbb{X} = (\mathbb{X}_{+}, \mathbb{X}_{-})$,  where $\mathbb{X}_{+}$: byte files from malware of a family,
  $\mathbb{X}_{-}$: byte files from non-malware or malware from other families, and  Test set, $\mathbb{Y}$}
  \KwOut{Labels for $\mathbb{Y}$ }
  $(\mathbb{S}_{+}, \mathbb{S}_{-}) \gets$ Convert to nucleotide sequence files $(\mathbb{X}_{+}, \mathbb{X}_{-})$\;
  $\mathbb{B}\gets$ Perform multiple sequence alignment and identify conserved blocks ($\mathbb{S}_{+}$)\; 
  \ForAll{blocks $B \in \mathbb{B}$}
    {
     $C_B\gets$ Get consensus sequence ($B$)\;
     \For{i = $1 \to length(B)$}
     {
        Calculate $ConservationScore(B,i,N)$
        where $N=A,C,G,T$\;
     }
    }
  \ForAll{sequences $Z \in (\mathbb{S}_{+}\cup \mathbb{S}_{-})$}
  {
    $F_Z$:= Feature vector of $Z$\;
    \ForAll{blocks $B \in \mathbb{B}$}
    {
    $S\gets$ Get alignments ($Z$, $C_B$)\;
    $AlignmentScore\gets $ Calculate alignment score $(S,B)$\;
    $AlignmentCount\gets $ Get alignment count $(S)$\;
    $F_Z\gets F_Z \cup (AlignmentScore, AlignmentCount) $\;
    }
  }
  $\mathcal{M}\gets$ Learn classification model  $(\mathbb{F},\mathbb{X})$ where $\mathbb{F}$: feature matrix\;   
 
    $\mathbb{T} \gets$ Convert to nucleotide sequence files ($\mathbb{Y}$)\;
    $\mathbb{L}$ : labels\;
 \ForAll{sequences $T \in \mathbb{T}$}
 {
    $F_T$:= Feature vector of $T$\;
    \ForAll{blocks $B \in \mathbb{B}$}
    {
    $S\gets$ Get alignments ($T$, $C_B$)\;
    $AlignmentScore\gets $ Calculate alignment score $(S,B)$\;
    $AlignmentCount\gets $ Get alignment count $(S)$\;
    $F_T\gets F_T \cup (AlignmentScore, AlignmentCount) $\;
    }
    $L \gets$ Predict $(\mathcal{M}, F_T)$\;
    $\mathbb{L}\gets \mathbb{L} \cup L$\;
 }
 \Return $\mathbb{L}$
\caption{\toolname}
\label{algo:malign}
\end{algorithm}

The main steps of our proposed method are shown in Algorithm~\ref{algo:malign} and an illustration is provided in Figure \ref{fig:overview}. We start with the given \emph{malware bytes files}, i.e., PE files~\citep{pefile} and convert them to \emph{malware \nucleicseq \ files}, i.e., sequences of A, C, G, and T, to make them compatible with multiple whole-genome alignment technique (SibeliaZ ~\citep{sibeliaz}). 
This outputs alignment blocks that are common among a number of these files. These alignment blocks are merged and thus \emph{\consensusseq}\ is constructed. In this step, \emph{conservation score} for each coordinate of the \consensusseq \ is also generated. This \consensusseq\ is aligned with each sample from a balanced training set with positive and negative samples with respect to the malware family of interest, and an \emph{alignment score} is calculated for every sample for each conserved block. These scores are then used as input to a machine learning model, which learns a classifier to distinguish between malware belonging to the family and malware from other families. In the inference step, to classify a new sample, the sequence is aligned with the consensus sequence, and alignment scores for the new instance are generated in a  similar way. The scores are passed into our classifier to classify the new sample. Each of these steps is described in more detail below.

\subsection{\textbf{Bytes file to Nucleotide Sequence file Conversion}}

First, the binary executable or bytes files are converted to \nucleicseq\ files containing sequences of A, C, G and T. The conversion is performed so that the existing whole-genome alignment tool can take the files as input. 
The conversion from the byte code to nucleotide sequence is done by converting each pair of bits to a nucleotide according to the following table: 

\begin{center}    
    \begin{tabular}{c c} 
        \hline
        $ 00 \rightarrow A $ \ \ & \ \  $ 01 \rightarrow C $\\ 
        $ 10 \rightarrow G $ \ \ & \ \ $ 11 \rightarrow T $\\
         [1ex]
        \hline
    \end{tabular}
\end{center}

Note that, similar conversions have been used by previous works, e.g. ~\cite{naidu2014further,naidu2016needleman, drew2016polymorphic}, etc., and were successful in showing that such nucleotide sequences still can preserve necessary functionalities from the byte files.

In some malware datasets such as the Kaggle Microsoft Malware Classification Challenge (Big 2015) dataset \citep{microsoftdataset2015}, the provided bytes files 
contain ``\texttt{??}'' and long stretches of  ``\texttt{00}'' in some cases which do not preserve any significant value or meaning. These are removed before the conversion to nucleotide sequences. Stretches of ``\texttt{00}'' longer than a  a threshold value of 32 were removed.

\subsection{\textbf{Multiple Alignment of Malware Nucleotide Sequences}}
The next step is to align the malware nucleotide sequences.  
In this paper, we adopt the multiple whole-genome alignment technique SibeliaZ \citep{sibeliaz}. 
It performs whole-genome alignment of multiple sequences and constructs locally co-linear blocks.
Figure \ref{fig:conservation score} illustrates alignment of three different malware \nucleicseq\ files from the same family. The sequences share blocks of similar sequences shown in dashed lines of the same color. They may also contain sequences unique to each sequence indicated by lines with different colors. 

During the block construction process:
the order of the shared blocks may differ in different sequences and the blocks may not be fully shared across all sequences. This helps \toolname\ to be robust against evasion attempts, such as reordering of instructions, injection of benign or random code at places, etc. In addition,  the shared blocks may not be fully conserved, i.e., there can be mismatches of characters to some extent which means minor alteration or modification to the code will not prevent the detection of blocks.
These properties of multiple whole-genome alignment improves the robustness of \toolname\ against many evasion attacks.

SibeliaZ first identifies the shared linear blocks and then performs multiple sequence alignment of locally co-linear blocks. The block coordinates are output in the GFF format \citep{gff_file} and the alignment is in the MAF format~\citep{multiple_align_format}. 
The multiple alignment format (MAF) is a format for describing multiple alignments in a way that is easy to parse and read. In our case, this format stores multiple alignment blocks at the byte code level among malware. 
We generate such an MAF file for each malware family using the training samples and identify the blocks of codes that are highly conserved across the malware family. 

\subsection{\textbf{Consensus Sequence and Score Generation}}
We process all sequences of the alignment blocks of the MAF file from the previous step and generate a new sequence for each block, which we will call the \emph{\consensusseq} -- motivated by \cite{consensus}. At first, we scan the length of all sequences and find the maximum one that will be the length of our \consensusseq. Then we traverse through the coordinates of every sequence and find the nucleotide of the highest occurrence for each coordinate. We put the most frequently occurring nucleotide at the corresponding index  of the \consensusseq, i.e., the characters of the \consensusseq s of the blocks are given by

$$C_{B,i}= \argmax_{N \in \{A,C,G,T\}}k_{B,i,N}$$ 
$\text{for } 1\leq i \leq l_B \text{ and } B\in \mathbb{B}$, where $k_{B,i,N}$ is the count of the character $N (= A, C, G, T)$ at the $i$-th position in block $B$ and $l_B$ is the length of block $B$.   

\begin{figure*}[ht!]
    \centering
    \includegraphics[width=0.75\textwidth]{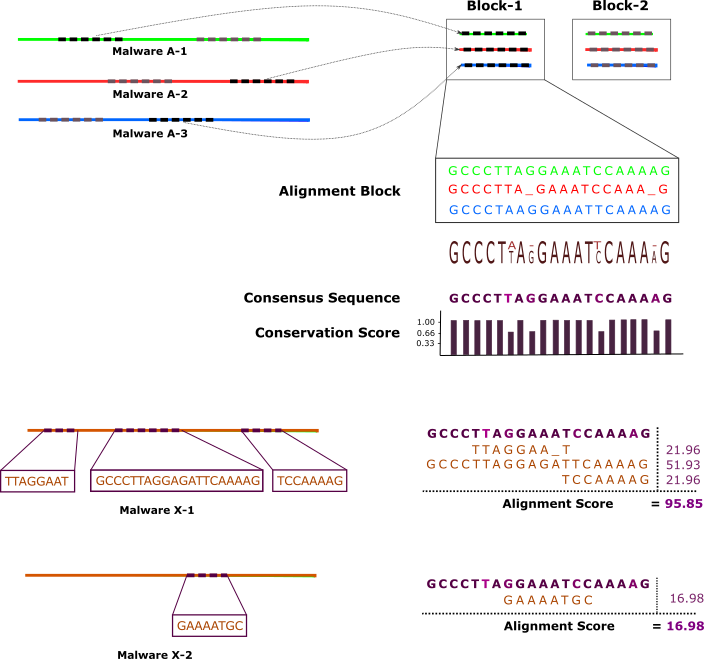}
    \caption{Details of \consensusseq, conservation score, and alignment score generation from alignment blocks. (1) Block 1 is generated from a shared common block among 3 Malwares from family A, where the code blocks can be in different orders. (2) Consensus Sequence and Conservation Score are generated from Alignment Block 1. (3) During inference time, for new Malware X-1 and X-2, they are aligned with Alignment Block 1, and scores are generated for every alignment. For example, Malware X-1 has a higher probability of belonging to Family A than Malware X-2.}
    \label{fig:conservation score}
\end{figure*}

In Figure \ref{fig:conservation score}, consensus sequence generation of an alignment block (Block-1) is shown in detail. Below the alignment block, the corresponding sequence logo is shown. The height of the individual letters in the sequence logo represents how common the corresponding letter is at that particular coordinate of the alignment.

We thus construct the \consensusseq \ by taking the letter (nucleotide) with the highest frequency for each coordinate.
Similarly, the \consensusseq s for all blocks are generated and are stored in a  file in FASTA~\citep{fastaformat} format with a unique id. These \consensusseq s are the conserved part of the malware family which can be considered as the signature or common pattern of that 
family. These files are later used to classify malware families.

In addition to the consensus sequence, we calculate \emph{conservation scores} for the blocks. In bioinformatics, conservation score is used during the evaluation of sites in a multiple sequence alignment, in order to identify residues critical for structure or function. This is calculated per base, indicating how many species in a given multiple alignment match at each locus. We adopted this concept in the malware domain where the conservation score can indicate the significance or importance of a code segment in a malware family. The responsible code segments of malware will have high conservation scores compared to the segments that are not frequent or conserved in malware files.

In Figure \ref{fig:conservation score}, the heights of the bars corresponding to conservation scores indicate the degree of conservation at the corresponding positions. For each coordinate, we store the score for each of the four nucleotides which is given by the  occurrence ratio of that nucleotide at that coordinate. So, \emph{conservation score} at the $i^{th}$ index of the alignment block $B$ for nucleotide $N$ is given by
$$\gamma_{B,i,N} =
\frac{k_{B,i,N}}{n_B} 
$$
where ${n_B}$ is the number of sequences that constructed block $B$.

For example, in Figure \ref{fig:conservation score},  Block-1 has 3 sequences in total. Since at the $1^{st}$ index, the block contains 3 $G$s, $$\gamma_{1, 1, G} = 3/3 = 1.00$$ 
Again at the $6^{th}$ index, the block contains 2 $T$s and 1 $A$. Therefore, 
$$\gamma_{1, 6, T} = 2/3 = 0.66 \text{ and } \gamma_{1, 6, A} = 1/3 = 0.33$$

\subsection{\textbf{Alignment with Consensus Sequences}}
Once the consensus sequences and the conservation scores are generated, we take a training set for each malware family. In the training set, the positive examples are samples from that malware family and the negative examples are malware from other families. All samples from the training set are aligned to the \consensusseq s of the corresponding family to get the aligned blocks for each sample. Using the previously generated conservation scores, we calculate new scores called \emph{alignment scores} for each block for all samples.

An example of alignment score calculation is shown in Figure \ref{fig:conservation score}. Malware X-1 and X-2 are positive and negative samples respectively. X-1 has three aligned sequences with the \consensusseq \ (shown in purple) whereas X-2 has only one. The sum of scores for all aligned sequences will be the score for the corresponding block of that sample. As an example, for the total score of sample X-1, we sum the scores of 3 aligned sequences, where each aligned sequence's score is the sum of the score of all coordinates. 

The aligned sequence score is then multiplied by the number of sequences that constructed the corresponding block since the higher the number of sequences that generated the block, the more conserved the sequence is across the malwares from that family. In Figure \ref{fig:conservation score}, 
adding the score for all coordinates of the first aligned sequence of the sample X-1, we get $7.32$. Since the corresponding \consensusseq \ was generated from 3 sequences, the final score for first aligned sequence will be $21.96 (= 7.32 \times 3)$.
Finally, the  total alignment score for the block was calculated by adding the scores of all 3 aligned sequences. 

In general, the \emph{alignmment score} of a sample $Z$ for \consensusseq \ $C_B$ of the block $B$ is given by
$$  
\alpha_{Z,B} =\sum _{s\in S}^{}\Biggl(\sum\limits _{i=1}^{length(s)} \gamma_{B,j,s_i}\Biggr) \times n_B\
$$
where,  $S$ is the set of sequences from sample $Z$ that got aligned to $C_B$, $s_i$ is the $i$-th nucleotide of the sequence $s$, and $j$ is the index of $C_B$ where $s_i$ was aligned.

Along with this score, we also store the total number of times the \consensusseq\ of a block gets aligned with the sample, i.e., \emph{alignment count} $\beta_{Z,B}=|S|$. 
Both the number of occurrences and the total alignment score for the \consensusseq\ of each block for a malware family are used as input for the subsequent classification, resulting in $2m$ features if a malware family has  $m$ aligned blocks.  

\subsection{\textbf{Classification}}
Finally, we train machine learning models for each malware family to classify malware. The scores and the number of alignments calculated as mentioned above are used as the features in our classifiers.

We experimented with a number of machine learning models including logistic regression, support vector machines (SVM), decision trees, and a simple deep learning model. 
Since the results did not vary significantly across models (see Table \ref{table:machine learning models result} in Results), we use logistic regression as our primary model for its better interpretability. We are going to refer this model as \textbf{\toolname\ (Logistic Regression)} in the rest of this paper. We also experimented with a simple deep-learning model to find if it gives any better results, which we are going to refer to as \textbf{\toolname\ (Deep Learning)}. However, such deep-learning models are not directly interpretable, and so, whenever we mention \toolname\ in this paper, we indicate the \toolname\ (Logistic Regression) version.

After the training phase, we get classifiers that can be used to classify new samples as shown in Figure \ref{fig:overview}. The scores of the training and the test samples are calculated in the same way. The scores for new samples are passed to the classifier for binary classification. 


%% file: files/results.tex
\label{sec:evaluation}

\subsection{\textbf{Datasets}}

\begin{table*}[h!]
    \centering
    {
    \begin{tabular}{c c c}
        \hline 
        Family Name & No of Train Samples & Type \\ [0.5ex] 
        \hline
        Ramnit & 1541 & Worm \\ 
        Lollipop & 2478 & Adware \\ 
        Kelihos\textunderscore ver3 & 2942 & Backdoor \\
        Vundo & 475 & Trojan \\
        Simda & 42 & Backdoor \\
        Tracur & 751 & TrojanDownloader \\
        Kelihos \textunderscore ver1 & 398 & Backdoor \\
        Obfuscator.ACY & 1228 & Any kind of obfuscated malware \\
        Gatak & 1013 & Backdoor \\ [1ex]
        \hline
    \end{tabular}
    }
    \caption{Malware Families in the Kaggle Microsoft Malware Classification Challenge Dataset}        
    \label{table:malware types}
\end{table*}

\subsubsection{The Kaggle Microsoft Malware Classification Challenge (Big 2015)}
The Kaggle Microsoft Malware Classification Challenge (Big 2015) ~\citep{microsoftdataset2015} aimed to organize a competition on malware family classification with a dataset consisting of 9 malware families (see Table \ref{table:malware types}).

This challenge simulates the data processed on over 160 million computers by Microsoft’s real-time anti-malware detection products inspecting over 700 million computers per month.
Microsoft provided almost half a terabyte of input training and classification input data when uncompressed. They included:
\begin{enumerate}
    \item \textit{Binary Files:} 10,868 training files containing the raw hexadecimal representation of the binary content of the malware.
    \item \textit{Assembly Files:} 10,868 training files containing data extracted by the Interactive Disassembler (IDA) tool. This information includes assembly command sequences, function calls, and more.
    \item \textit{Training Labels:} Each training file name is an MD5 hash of the actual program. Each MD5 hash and the malware class it maps to are stored in the training label file.
\end{enumerate}

From this, we constructed 9 balanced datasets for binary classification consisting of an equal number of positive and negative samples for each of the 9 malware families. In each dataset, the positive examples are all the samples from the corresponding family and the negative examples were chosen by randomly sampling from the 8 other families in the dataset. Then 20\% of each dataset was set aside as the test sets while the remaining 80\% was used as the training sets. For the deep learning approaches that require hyper-parameter selection, the 80\% was further split into training (60\%) and validation sets (20\%). \\



\subsubsection{Microsoft Machine Learning Security Evasion  Competition (2020)  (MLSEC) Dataset}~\\
While the Kaggle Microsoft Malware Classification Challenge (Big 2015) dataset is a large and widely studied one, often new malware families only have a few samples - especially during their initial stages. Moreover, they might not always accurately represent real-world scenarios and  in-the-wild practice. Therefore to assess the performance of our method on in-the-wild datasets with a small number of instances per family, we used the Microsoft Machine Learning Security Evasion  Competition (2020) \citep{mlsec2020} dataset too. \footnote{The dataset may not remain public by Microsoft by the time of publication, but we can provide the dataset upon request.}

In this competition, the defenders' challenge was to create a solution model that can defend against evasive variants created by the attackers. The defenders were provided a `Defender Challenge' dataset with 49 original malware and their evasive variants created by real malware attackers.
Each malware contains a different number of evasive variants varying from 5 to 20. On average, malware has 12 variants in this dataset. We chose this dataset because --
\begin{inparaenum}[i)]
    \item it mirrors different types of real-life attacks, and 
    \item it has a limited number of samples which can be a good representation of new emerging malware.
\end{inparaenum}

Similarly to the Kaggle Microsoft Malware dataset, 49 different datasets were created for each family, which were then split into training, validation, and test sets. During the split of this dataset, we always kept the original sample in the train set and the variants in the test set for each family, so that the test set can be considered as an evolution of the training set.

\subsection{\textbf{Evaluation of Machine Learning Algorithms}}

First, we assess the performance of various machine learning approaches on the Kaggle Microsoft Malware (Big 2015) dataset. The binary files of this dataset were converted to \nucleicseq \ files and labeled using `Training Labels’ data provided by Microsoft. Then we generated common alignment blocks using SibeliaZ and constructed the \consensusseq s as discussed in  Section \ref{sec:method}. We generated the conservation scores for each \consensusseq s using the frequency of nucleotides which were then used as features for the machine learning models.

We experimented with logistic regression, decision trees, and support vector machines (SVM). 
Table \ref{table:machine learning models result} shows the test accuracy for 80\%-20\% train-test split on the Kaggle Microsoft Malware Classification Challenge (Big 2015) Dataset. 
We observe that the models show similar performances in terms of accuracy. So, we selected \toolname\ (Logistic Regression) for future experiments because of its simplicity and interpretability. For hyper-parameters details and results on the train-set, readers can refer to \ref{app_sec:model} and \ref{app_sec:result}.

\begin{table*}[h!]
    \centering
    {
    \begin{tabular}{c |c c c } 
        \hline 
        \multirow{2}{*}{Family Name} & \multicolumn{3}{c}{Test Accuracy} \\
        {} & Logistic regression & Decision tree & SVM \\
        \hline
        Ramnit & 99.64 & 99.82 & 99.64 \\ 
        Kelihos\textunderscore ver3 & 99.27 & 99.27 & 99.27 \\
        Vundo & 97.4 & 97.4 & 97.4 \\
        Simda &  84.62 & 84.62 & 84.62 \\
        Tracur & 97.3 & 94.6 & 96.3 \\
        Kelihos \textunderscore ver1  & 96.7 & 98.9 & 98.9 \\
        Obfuscator.ACY & 95.9 & 92.7 & 96 \\
        Gatak  & 96.2 & 96.2 & 96.2 \\
        \hline
        Overall & 97.99 & 97.42 & 98.02 \\
        [0.5ex]
        \hline
    \end{tabular}
    }
    \caption{Test Accuracy on the Kaggle Dataset for different Machine Learning Models}        
    \label{table:machine learning models result}
\end{table*}

\subsection{\textbf{Comparison with Existing Approaches}}

Next, we compare the performance of \toolname\ with some of the state-of-the-art approaches that can take raw-byte, MalConv~\citep{malconv} (a deep-learning based approach using raw byte sequence), Feature-Fusion~\citep{ahmadi2016novel} (a feature extraction, selection and fusion based approach using byte and assembly files), and M-CNN~\citep{kalash2018malware} (a CNN based approach relying on conversion to images) on the Kaggle Microsoft Malware (Big 2015) dataset. It is worth noting that models with multiclass loss as low as 0.00283 have been reported for this specific dataset. However, we compare with MalConv and M-CNN, as they have been successfully applied to many different datasets, and to our knowledge, the Feature-Fusion method has even better accuracy (multiclass logloss = 0.00128) than the winning team of the Kaggle competition.  
Besides simple ML models, we also implement a deep-learning based model on top of the alignment scores calculated by \toolname, which we refer to as \toolname\ (Deep Learning).
The architectures of the \toolname\ (Deep Learning), MalConv and M-CNN are shown in Figure~\ref{fig: architectures} (see appendix).

\begin{table*}[h!]
    \centering
    \resizebox{0.8\textwidth}{!}{
    \begin{tabular}{c |c c| c c c } 
        \hline 
        \multirow{2}{*}{Family Name} & \toolname & \toolname & \multirow{2}{*}{Feature-Fusion} & \multirow{2}{*}{MalConv} & \multirow{2}{*}{M-CNN} \\ 
        & (Logistic Regression) & (Deep Learning) & & & \\ [0.5ex] 
        \hline
        Ramnit & 99.64 & 99.46 & 98.7 & 95.66 & 88.44 \\ 
        Kelihos\textunderscore ver3 & 99.27 & 99.82 & 99.18 & 100 & 99.72 \\
        Vundo & 97.4 & 98.70 & 95.79 & 94.89 & 97.21 \\
        Simda & 84.62 & 84.62 & 76.47 & 52.94 & 62.5 \\
        Tracur & 97.3 & 98.2 & 98.34 & 93.91 & 94.55 \\
        Kelihos \textunderscore ver1 & 96.7 & 95.7 & 98.75 & 96.08 & 94.67 \\
        Obfuscator.ACY & 95.9 & 96.39 & 98.98 & 94.42 & 91.74 \\
        Gatak & 96.2 & 98.37 & 98.03 & 98.67 & 88.68 \\
        \hline
        Overall & 97.99 & \textbf{98.59} & 98.52 & 96.95 & 94.10 \\
        [0.5ex]
        \hline
    \end{tabular}
    }
    \caption{Test Accuracy of all Models on the Kaggle Dataset}        
    \label{table:microsoft classification test accuracy for different models}
\end{table*}

\begin{table}[h]
    \centering
    \resizebox{\columnwidth}{!}{
    \begin{tabular}{c c c c c}
        \hline 
        Model & Recall & Specificity & Precision & F1 Score \\ [0.5ex] 
        \hline
        \toolname & \multirow{2}{*}{96.66} & \multirow{2}{*}{\textbf{98.60}} & \multirow{2}{*}{\textbf{98.67}} & \multirow{2}{*}{97.61}\\
        (Logistic Regression) & & & & \\
        \toolname & \multirow{2}{*}{\textbf{99.00}} & \multirow{2}{*}{98.18} & \multirow{2}{*}{98.19} & \multirow{2}{*}{\textbf{98.59}}\\
        (Deep Learning) & & & & \\
        \hline
        MalConv & 97.37 & 96.48 & 97.36 & 97.3 \\
        M-CNN & 93.44 & 94.85 & 94.59 & 93.94 \\ [1ex]
        \hline
    \end{tabular}
    }
    \caption{Performance of Models on the Kaggle Dataset for other Evaluation Metrics}
    \label{table:microsoft performance test-set}
\end{table}

The test and training accuracy of \toolname\ with logistic regression and deep learning along with those of Feature-Fusion, MalConv, and M-CNN are shown in Tables  \ref{table:microsoft classification test accuracy for different models} and ~\ref{table:microsoft classification train accuracy for different models} (appendix).
Table \ref{table:microsoft classification test accuracy for different models} shows that \toolname \ (Deep Learning) has the best accuracy on the test set than other approaches. 
\toolname \ (Logistic Regression) and (Deep Learning) have 97.99\% and 98.59\% accuracy which is close to the Kaggle winning team's performance.

We compared \toolname \ with other models with respect to recall, specificity, precision, and F1 score. 
Table \ref{table:microsoft performance test-set} represents the overall performance of models taking the weighted average for all types (detailed result for each family in Table \ref{table:microsoft performane test-set (all)} in the appendix). It shows that \toolname \ has a more balanced performance than others. Moreover, it has higher precision which indicates that it has a comparatively low false positive rate -- an indicator of a good malware detection model in commercial AV systems.


\subsection{\textbf{Applicability with Limited Amount of Data and Features}}

\begin{figure*}[!htb]
\centering
\subfloat[\label{fig:train mlsec radar}Radar Chart showing Accuracy on MLSec-Train Dataset]{\includegraphics[width = 0.5\textwidth]{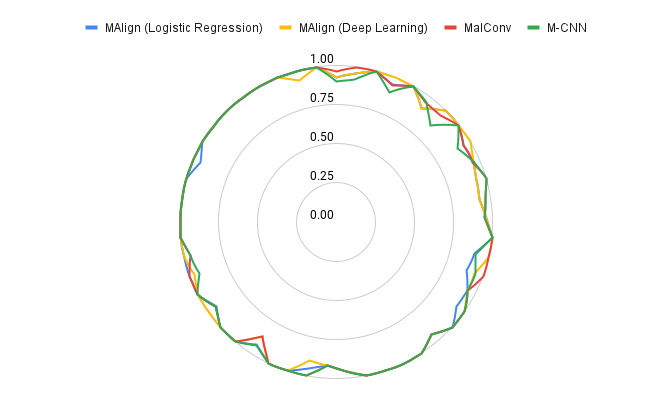}}
\hfill
\subfloat[\label{fig:test mlsec radar}Radar Chart showing Accuracy on MLSec-Test Dataset]{\includegraphics[width = 0.5\textwidth]{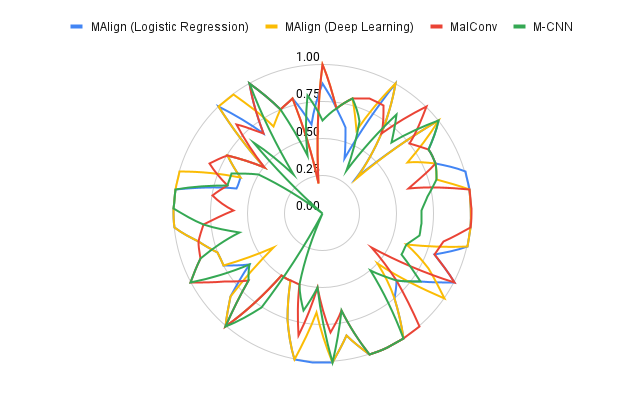}}
\caption{Radar Chart showing Accuracy of all Models along the perimeter on each of the individual 49 malware types in the MLSec Dataset}
\label{fig:mlsec radar}
\end{figure*}

Although deep learning based approaches 
have been widely applied for malware classification and detection, they require extensive amounts of data for training and tend to overfit in the absence of that.  Table \ref{table:microsoft classification test accuracy for different models}  shows that Feature-Fusion, MalConv, and M-CNN perform arguably well for most malware families. However, we observe that, for Type 5 (Simda), which has only 42 samples, the test accuracies of Feature-Fusion, MalConv, and M-CNN are respectively 76.47\%, 52.94\% and 62.5\%  while \toolname \ has 84.62\% test accuracy. 
A similar observation can be made for Type 4 (Vundo) which has the second smallest number of training samples. 

Moreover, \toolname\ only needs the raw byte sequence whereas the typical methods like Feature-Fusion need other information or features, e.g., assembly file, address for byte sequence, section information from PE, etc. In addition, \toolname\ can take any arbitrary length of input sequence whereas E2E deep learning based approach like MalConv, M-CNN suffers from limited input-length constraint.

\begin{table}[h]
    \centering
    \resizebox{\columnwidth}{!}{
    \begin{tabular}{c c c}
        \hline 
        Model & Train Accuracy & Test Accuracy \\ [0.5ex] 
        \hline
        \toolname\ (Logistic Regression) & 98.18 & \textbf{80.42} \\ 
        \toolname\ (Deep Learning) & 97.72 & 80.00 \\ 
        [0.5ex]
        \hline
        Feature-Fusion & 98.02 & 60.94 \\
        MalConv & \textbf{98.24} & 79.22 \\
        M-CNN & 96.99 & 71.09 \\ [1ex]
        \hline
    \end{tabular}
    }
    \caption{Performance of Models on the MLSEC dataset for train-test split}
    \label{table:MLSEC performance train-test}
\end{table}

We also evaluated the performances of the methods on the MLSEC dataset (Microsoft Machine Learning Security Evasion  Competition (2020) \citep{mlsec2020}). Since it contains a limited number of variants created in almost real-time, this can be used to identify how our method works on new variants of a malware when only a limited number of samples are available. Because of the limited number of instances in this dataset, the validation set is very small for some types. Table~\ref{table:MLSEC performance train-test} shows the weighted result of all models for an 80\%-20\% train-test split on all 49 malware families. Additionally, we evaluated another set of results for 60\%-20\%-20\% train-validation-test split just for deep learning models (see Figure \ref{table:MLSEC performance train-val-test} in appendix).
We observe that \toolname\ outperforms the other models on the MLSEC dataset regardless of the splitting.

Furthermore, Figures \ref{fig:train mlsec radar} and \ref{fig:test mlsec radar} provide radar charts that visually illustrate the training and test accuracy for all 49 types individually.
From these figures, we observe that on the train set, all approaches have consistent performance, but on the test set M-CNN and MalConv are relatively inconsistent\footnotemark{}. For example, types 29 and 43 have 10 and 5 available variants respectively, and M-CNN's test accuracy is 0 on both, indicating a limitation in its performance on these specific variants.
\footnotetext{The more circular the perimeter line is, the more consistent the performance is.}

For further analysis, we experimented with dropping the train-set size of the Tracur family of the Kaggle dataset to half and evaluated all models.
We observed that \toolname\ has the least accuracy drop compared to others (see Table \ref{table:test_acc_drop}).
Again, this proves the advantage of our method when a malware family does not have a considerably large sample size, which is very common in the case of emerging malware families. This was one of the motivations for adopting sequence alignment since its capability of explicit identification of critical code blocks still sustains the scarcity of data.



\begin{table}[h]
    \centering
    \resizebox{\columnwidth}{!}{
    \begin{tabular}{c c c } 
        \hline 
        \multirow{2}{*}{Model} & Accuracy & Accuracy   \\
         & Drop & Change (\%) $\downarrow$ \\
        \hline
        \toolname(Logistic Regression) & $97.3\% \rightarrow 92.56\%$ & $\mathbf{4.87\%}$ \\ 
        \toolname(Deep Learning) & $98.2\% \rightarrow 92.67\%$ & $5.53\%$ \\
        \hline
        MalConv & $93.91\% \rightarrow 85.67\%$ & $8.77\%$ \\
        m-CNN & $94.55\% \rightarrow 83.22\%$ & $11.98\%$ \\
        [0.5ex]
        \hline
    \end{tabular}
    }
    \caption{Test Accuracy Drop for dropping the train-set size of Tracur family to half}        
    \label{table:test_acc_drop}
\end{table}

%% file: files/explainability.tex
\label{sec:explainability}

An important feature of \toolname\ which makes it different from other classifiers is its interpretability, and its capability of deriving suspicious code blocks in malware families through a simple \emph{backtracking} process. Upon analysis, the `maliciousness' of these suspicious code blocks can be determined. The \emph{backtracking} process is as follows:

\begin{enumerate}[(i)]
\itemsep0em
    \item Finding the blocks that are assigned high weights by the logistic regression model.
    \item Selecting the blocks (found from step (i)) that are highly conserved.
    \item Processing the multiple alignment file (MAF) to find out the sequences and their indices that constructed the blocks.
    \item Locating the code fragments in binary and assembly files (using a disassembler) corresponding to the sequences found in step (iii).
\end{enumerate}

We evaluated the interpretability of our \toolname\ by following the above-mentioned backtracking process on multiple families from the Microsoft Kaggle Dataset~\footnote{The MD5 hashes mentioned in this section are provided by the Kaggle Microsoft Dataset.}. We discuss some of the interesting findings below (more in \ref{app_sec:explain}).

\subsection{Kelihos\_ver1 Backdoor}

\begin{figure*}[!htb]
    \centering
    \includegraphics[width=0.9\textwidth]{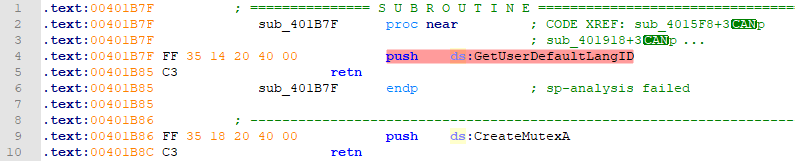}
    \caption{Kelihos (version 1) Backdoor tracking user geological position for botnet connection \\(MD5 hash = aOCU2V7b0RkgQt9LflYF)}
    \label{fig:kelihos1_lang}
\end{figure*}

The Kelihos backdoor allows the affected system to be part of the Kelihos botnet. It is then typically used to send spam email messages, steal information, etc.
~\citep{fsecure_kelihos1}


In our analysis of this specific backdoor malware, interestingly, one of our consensus sequences got mapped to 119 samples. While looking into the assembly code of this signature, we found that all of them are calling \texttt{\textbf{GetUserDefaultlangID}} function (see Figure \ref{fig:kelihos1_lang}). Usually, this function is called by malware to track geological position and trigger a regional attack~\citep{malware_func_sheet}.

Furthermore, we found more code blocks having suspicious Windows function calls from \texttt{\textbf{IEAdvpack.dll}}
which gets abused by malware authors very often. From Figure \ref{fig:kelihos1_api_calls} in Appendix, we want to emphasize that -- the order of these function calls is different in malwares, e.g., \texttt{IsNTAdmin} function is declared before \texttt{RebootCheckOnInstall} in \ref{fig:kelihos1_1} and \ref{fig:kelihos1_2}, but not in \ref{fig:kelihos1_3}. However,  \toolname\ can successfully detect them due to its flexibility in alignment. The common pattern in these malwares from Figure \ref{fig:kelihos1_api_calls} is that -- they are checking for admin privileges by \texttt{IsNTAdmin}, calling suspicious function \texttt{RebootCheckOnInstall} (also found in previous reports from Falcon Sandbox of~\cite{falcon_kelihos1_1,falcon_kelihos1_2}), and deleting files sometimes using \texttt{DelNodeRunDLL32}. 

\subsection{Tracur Trojan-Downloader}

Trojan-Downloader:W32/Tracur.J is a malware that is known to identify a malicious DLL file that installs a plug-in for Internet Explorer and/or Mozilla Firefox web browsers and then redirects searches to an unsolicited website~\citep{fsecure_tracur}.

\begin{figure*}[ht]
    \centering
    \includegraphics[width=0.7\textwidth]{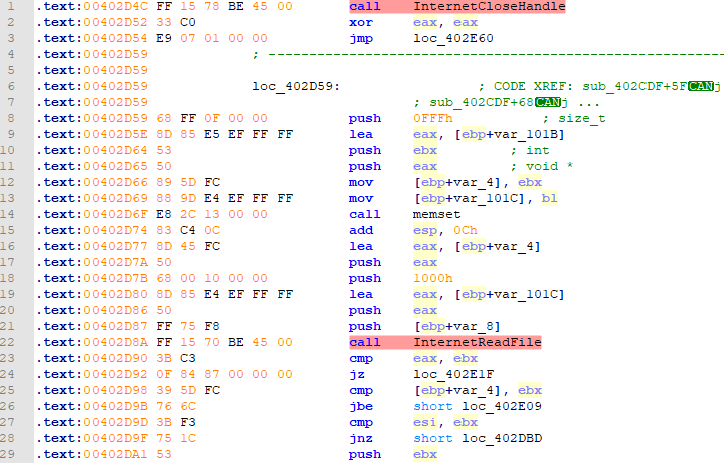}
    \caption{Tracur Trojan-Downloader trying to read file from internet (MD5 hash = gEZCMz90lrmI8cx37FNT)}
    \label{fig:tracur_read_file}
\end{figure*}

Consequently, malware authors write malware from \emph{Tracur} family in such a way that it can get connected to the internet and download or upload the desired file (close to banking trojan ~\citep{ioactive2012reversal}). Our consensus sequence successfully detected code segments for such activities. Figure \ref{fig:tracur_read_file} is an example of such malware which is calling functions like -- \texttt{InternetReadFile, InternetCloseHandle}. This clearly reveals that \toolname\ is capable of capturing such sequential function calling that is suspicious.

\subsection{Obfuscator.ACY}

Malware with vastly different purposes often ``obfuscate" i.e. hide its purpose so that security software does not detect it~\citep{microsoft_obfuscator}.
While different obfuscator tools can obfuscate or protect the malicious code in different ways, some packers like \textit{Themida} keep one function -- \texttt{\textbf{TlsSetValue}} ~\citep{obfuscate_tlssetvalue} so that it can unpack it later. This function is used to set the value of a pointer allocated for a TLS variable and has been heavily used by malware authors for many different attacks ~\citep{hellal2016minimal, poudyal2021analysis}, e.g., stack-smashing attack~\citep{nebenzahl2006install}. 

\begin{figure*}[ht]
    \centering
    \includegraphics[width=0.9\textwidth]{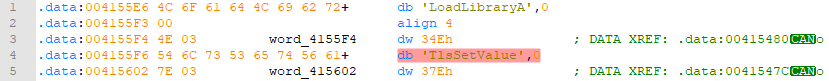}
    \caption{Detection of non-removable \texttt{TlsSetValue} function by Obfuscator tool (MD5 hash = IjoyeRsfBXW3bOUDMArL)}
    \label{fig:obfuscate_tlssetval}
\end{figure*}

In our analysis, we found consensus sequences getting matched with code blocks that contain \texttt{\textbf{TlsSetValue}}. Figure \ref{fig:obfuscate_tlssetval} shows the code block from one of such consensus sequences which was present in 39 samples from \emph{Obfuscator.ACY} family.

\subsection{Vundo Trojan}

Vundo is malware that downloads and displays pop-up advertisements of rogue software. It can download and install other malware and there are variants that collect information such as IP address, MAC address, Windows and browser versions, etc.\ and send it to attackers~\citep{fireeyevundo}.


While analyzing random malware executables from \emph{Vundo} family, surprisingly, we found some consensus sequences that were mapped to  binary sequences from different sections. To evaluate whether they are just some spurious features or not, we examined their assembly code and interestingly found that there is obfuscated code (from \texttt{DllEntryPoint} function) put into the \texttt{`data'} section instead of \texttt{`text'} section. It might be the case that -- one malware is loading another in its \texttt{data} section. However, such patterns got identified by our \toolname\ since their binary representations are similar (see Figures \ref{fig:malware obfuscation 4} and \ref{fig:vundo obfuscation 5} in Appendix).

\subsection{Simda Backdoor}

Backdoor:W32/Simda is a large malware family that allows attackers to remotely control machines they are installed in to steal personal or system data, take screenshots, download additional files onto the system, etc.  
~\citep{fsecure_simda}


\begin{figure*}[ht]
    \centering
    \includegraphics[width=\textwidth]{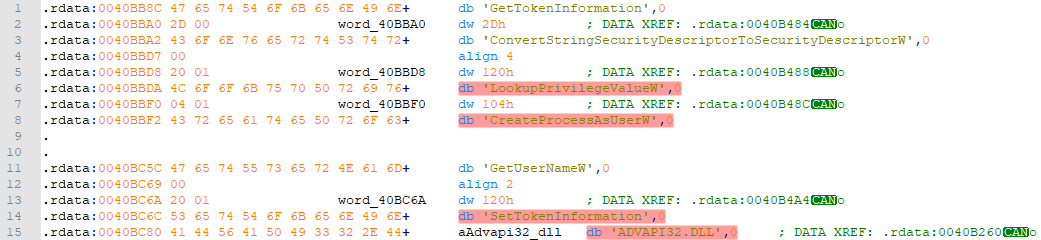}
    \caption{Suspicious function calls that matched with our consensus sequence for Simda Backdoor \\(MD5 hash = hrguW0CqMt2mJdfvX8IB)}
    \label{fig:simda_calls}
\end{figure*}

Among the \toolname\ generated consensus sequences for \emph{Simda} family, we randomly selected some of them and found matches to some suspicious function calls. For example, two samples had exactly the same code where they were declaring suspicious functions like -- \texttt{LookUpPrivilegeValueW, CreateProcessAsUserW, SetTokenInformation,} etc.
Then, they are using \texttt{\textbf{ADVAPI32.DLL}} through which they can access core windows components such as the service manager and registry (see Figure \ref{fig:simda_calls}). It might be the case that the backdoor malware is checking for the privilege to perform privileged operations, or checking if the privilege escalation exploit should be executed later.

%% file: files/robustness.tex
\label{sec:robustness}

A major issue with static malware detection approaches is - they can be evaded by malware authors in multiple ways, such as adding content without altering malware functionalities. There are several ways to find or generate this content, such as gradient-based patch ~\citep{kreuk2018deceiving, kolosnjaji2018adversarial, suciu2019exploring}. Since \toolname\  relies on finding conserved blocks critical to the malware families for classification through sequence alignment and score calculation, and an interpretable logistic regression model, it should in principle to be inherently robust to such attacks.

\subsection{\textbf{Theoretical Robustness}}
To formally show the difficulty of such attacks, as in~\cite{adv_ml_lorenzo, guo2018defending}, we mathematically prove that inverting our whole genome alignment process is an NP-Hard problem.

\input{files/robust_proof}
While we acknowledge that -- theoretical robustness does not ensure practical robustness, this property of \toolname\ shows that the model is robust to certain kinds of attack. 


\subsection{\textbf{Empirical Robustness}}
We also investigated the empirical robustness of our method compared to the other conventional malware detection techniques. 
We generated adversarial malwares using gradient-based patch attack of~\cite{kolosnjaji2018adversarial}. It creates adversarial samples just by modifying (or padding) approximately 1.25\% of the total size of malware which can successfully evade\footnotemark{} the MalConv model with a high percentage. The evasion rate increases with the percentage of modification on malware samples. We experimented with their implementation ~\citep{gradient_attack_github} on some of the types in the MLSEC dataset.

\footnotetext{If a malware gets detected by a model but its adversarial version gets undetected, then we call it a `successful evasion'.}

\vspace{-0mm}
\begin{table}[h!]
    \centering
    \resizebox{\columnwidth}{!}{
    \begin{tabular}{c |c c |c| c}
        \hline 
        & \multicolumn{3}{c|}{MalConv} & \toolname\\
        \hline
        \multirow{2}{*}{Type} & \multicolumn{2}{c|}{Evaded/Total} & \multirow{2}{*}{Evasion Rate} & \multirow{2}{*}{Evasion Rate}\\ 
        {} & Train set & Test set & {} \\[0.5ex] 
        \hline
        1 & 5/15 & 2/4 & 36.84\% & 0.00\% \\ 
        11 & 1/9 & 1/2 & 18.18\% & 0.00\% \\ 
        12 & 5/13 & 2/4 & 41.18\% & 0.00\% \\
        28 & 5/8 & 2/2 & 70\% & 0.00\% \\ 
        45 & 4/5 & 2/2 & 85.71\% & 0.00\% \\[1ex]
        \hline
    \end{tabular}
    }
    \caption{Evasion Rate on MalConv and \toolname\ using Gradient-based patch attack for some types in the MLSEC Dataset}
    \label{table:Gradient attack}
\end{table}

We then applied these evasive samples on \toolname\, and all of them were successfully detected with almost 100\% prediction confidence (see Table \ref{table:Gradient attack}). 
The probable reason for such a difference in the evasion rate is -- gradient-based patch attack~\citep{kolosnjaji2018adversarial} appends random bytes at the end of a malware file to evade the classifier, and since the alignment mechanism of \toolname\ is robust to insertions, its decision remains unaltered. Moreover, due to the non-invertible property of \toolname\ (proved in subsection \ref{app_sec:robust_proof}), the attacker cannot generate any such random content with gradient approach to evade it.

It is worth mentioning that -- no static analyzer can provide robustness to all type of attacks (similarly, no attack can evade all defense models). So, as security researchers, our goal is to make the bar higher for attackers. Since \toolname\ finds conserved code blocks in the presence of substitutions, insertions, and deletions of instructions, modifying those code segments sufficiently while preserving the malware semantics will be more challenging than other used models. However, we have discussed more about evasion techniques for \toolname\ in the \ref{app_sec:robust_discussion}.



%% file: files/robust_proof.tex
\subsubsection{Proof} \label{app_sec:robust_proof}
\begin{table}[h!]
    \centering
    \resizebox{0.8\columnwidth}{!}{
    \begin{tabular}{c | c}
        \hline 
        Symbol & Description \\ [0.5ex] 
        \hline
        Z & Problem space (i.e., input space) \\ 
        X & Feature Space \\ 
        Y & Label Space \\
        $\varphi$ & Feature mapping function, $\varphi : Z \rightarrow X$ \\
        g & Classifier, $g : X \rightarrow Y$ \\ [1ex]
        \hline
    \end{tabular}
    }
    \caption{Symbols with Description}
    \label{table:symbols}
\end{table}

We consider problem space $Z$ that contains malware files $z$ each having a ground truth label $y \in Y$. The feature mapping function $\varphi$, i.e., the alignment tool, score generation algorithm, etc, maps the input $z$ to a feature vector x; $\varphi(z)=x$. The machine learning classifier predicts a label $y$ for $x$; $g(x)=y$.

\begin{figure}[h]
    \centering
    \includegraphics[width=\columnwidth]{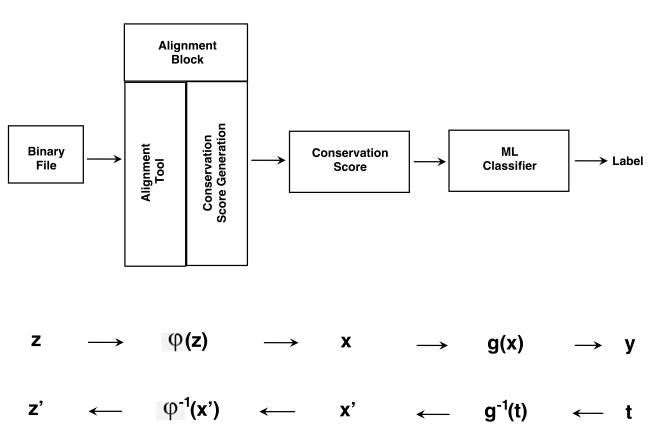}
    \caption{Flow of our method and the backward flow for attacks}
    \label{fig:robustness proof}
\end{figure}

Let's consider that all training and classifying algorithm is disclosed to the attacker.
The attacker's goal would be to perturb the input $z$ to $z'$ in such a way that our model misclassifies it as label $t$, where $t \neq y$. Since the machine learning classifier is a differentiable model, it is possible to find perturbed feature vector $x'$ through the gradient attack, such that, $g^{-1}(t)=x'$. The next job of the attacker would be to find an inverse feature mapping function, such that $\varphi^{-1}(x')=z'$. Now to block this backward flow, the feature mapping function $\varphi$ needs to be non-invertible and non-differentiable. 

\subsubsection*{Hardness of Invertibility}
Let, $B=\{B_{1}, B_{2}, B_{3}, ... , B_{m}\}$ denote the set of $m$ alignment blocks of a malware family where each element $B_{i}$ is a string of nucleotides.
The $i^{th}$ alignment block $B_{i}$ is represented by a conservation score matrix $A_{i} \in [0,1]^{5 \times n}$ which stores the conservation scores for each nucleotide, where $n$ is the length of the block. So, each row of the score matrix $A$ will store the scores for A, C, G, T and gap, respectively.

Let, $S_{i}=\{ S^{1}_{i}, S^{2}_{i}, S^{3}_{i}, ... , S^{c}_{i} \} $ denote the set of sequences from perturbed malware file $z'$ that got aligned with alignment block $B_{i}$ for $c$ times. Let's assume, without loss of generality, that only one sequence from input $z'$ is aligned with alignment block $B_{i}$ and so only one alignment score $\alpha_{z, B_{i}}$ is generated. 

So, to prove the hardness of invertibility, it suffices to prove that reconstructing the alignment sequence $S_{i}$ from the conservation score matrix $A_{i}$ and the alignment score $\alpha_{z, B_{i}}$ is hard i.e. the following problem is NP-hard. 

\textit{Alignment Inversion Problem:} 
Given a conservation score matrix $A \in [0,1]^{5 \times n}$ and a target alignment score $\alpha$, the alignment inversion problem is to find $y_{i,j}\in \{ 0,1 \}^{5 \times n}$ such that
$$\sum_{i=1}^{5}\sum_{j=1}^{n} y_{i,j} A_{i,j}=\alpha$$
Here the conservation scores for each column in matrix $A$ is normalized, i.e., $\sum_{i=1}^{n} A_{i,j} = 1$ for all $j$.

In order to show that the Alignment Inversion Problem is NP-hard, we reduce a well-known NP-complete problem, the subset sum problem, to our problem.

\textit{Subset Sum Problem:} The subset sum problem (SSP) is, given a set of integers and a target sum $T$, to decide whether any subset of the integers sum to $T$. The problem is NP-complete even when the numbers are restricted to positive integers. So, SSP refers to finding variables
$( x_{1}, x_{2}, x_{3}, ... , x_{n} ) \in \{ 0,1 \}^{n} $, given positive integers $( a_{1}, a_{2}, a_{3}, ... , a_{n}) \in \mathbb{N}^{n} $ and $T$, such that 
$$ \sum_{j=1}^{n} x_{j} a_{j} = T $$

Now, given an SSP instance, we generate an instance of the alignment inversion problem by setting
$$A_{1,j} = \frac{a_{j}}{N},\ A_{5,j} = 1 - \frac{a_{j}}{N},\ A_{i,j} = 0 \mbox{ for }i=\{2, 3, 4 \}$$ 
$$\mbox{ and }\alpha = \frac{T}{N}$$

Here $N$ is a positive integer chosen to make sure
that i) $0\leq A_{i,j} \leq 1$ for all $i,j$, and ii) $A_{5,j}= 1 - \frac{a_{j}}{N}\geq \alpha$ for all $j$. This can be done by setting $$N>a_{max}+T$$ where $a_{max}=\max(a_1, a_2, \dots, a_n)$.

Since $A_{5,j}\geq \alpha$ for all $j$, none of the scores for gaps can be included in the target alignment score $\alpha$, and $A_{i,j}$ for $i=2,3,4$ cannot contribute towards $\alpha$ as they are set to 0. Therefore, given the solution to the instance of one problem, we can obtain the solution to the other by setting $x_{j}=A_{1,j}$ for all $j$, because of the following: 
$$ \sum_{j=1}^{n} x_{j} a_{j} = T \ \ \Longleftrightarrow \ \ \frac{1}{N}\ \sum_{j=1}^{n} x_{j} a_{j} = \frac{T}{N} \ \ \Longleftrightarrow \ \ \sum_{j=1}^{n} y_{1,j} A_{1,j} = \alpha $$

Since SSP is a NP-hard problem and polynomial-time reducable to our problem, our problem is a NP-hard problem too. In our original problem, we may have to consider multiple nucleotides in most cases which is a harder version of this instance. So, it is trivial to prove that, our original problem is a NP-hard problem too.



%% file: files/app_runtime.tex
\section{Running Time} \label{app_sec:runtime}

For our experiments, we used a CPU of 40GB memory and GeForce RTX 3090. \toolname\ only requires the CPU, whereas we had to use the GPU for the deep learning based methods.
\subsection{\textbf{Mean-Time-To-Detect (MTTD)}}

\begin{table}[h!]
    \centering
    \begin{tabular}{c c c}
        \hline 
        Method & MTTD & Training Time \\ [0.5ex] 
        \hline
        \toolname & 0.0112584 s & 18hr 4min \\ 
        MalConv & 1.2461524 s & 18hr 13min \\ 
        M-CNN & 0.8706633 s & 14hr 26min \\ [1ex]
        \hline
    \end{tabular}
    \caption{MTTD and training time for different methods on the MLSEC dataset} \label{table:model runtime}
\end{table}

We evaluated the MTTD for all methods by considering the time they take in their inference stage. We ran all methods on the test set of the MLSEC dataset and measured the total time for the classifier to make the decisions. Then we divided this time by the total number of samples in the test set to find the MTTD. 

Table \ref{table:model runtime} shows the MTTD for each method. Since MalConv and M-CNN are deep learning models unlike \toolname\ (Logistic Regression), they take comparatively more time to classify a sample. Notably, we did not include the feature extraction time here, for which \toolname\ takes $15.9498$ seconds on average. In this feature extraction step, the graph construction step by TwoPaCo~\citep{minkin2017twopaco} takes up to $12$ seconds which is $75.24\%$ of the whole process. Since we do not have any control over the implementation of TwoPaCo, we cannot speed this process up anymore. However, once the signatures are found in \toolname, they can be trimmed if needed (based on how conserved they are) and then the alignment of the sequence of the new instance and the signature will be faster.

\subsection{\textbf{Training Time}}

We also evaluated the total training time for each method on the MLSEC dataset. From Table \ref{table:model runtime}, we can see that the training time is not vastly different from other methods. Here, we have included the time for alignment and feature extraction too.

~\cite{ahmadi2016novel}'s Feature-Fusion method has not been included in Table \ref{table:model runtime} because for the MLSEC dataset, partial features (only features from byte code) were considered. But we can get an estimation of its running time on the Kaggle Microsoft dataset from the original paper~\citep{ahmadi2016novel}. For example, it takes almost 17 hours and 15 minutes to extract only the `REG' feature from all samples, and it has to extract 14 features in total.

%% file: files/discussion.tex
In this work, we proposed \toolname\ for malware family classification from raw executables -- distinct from the prior works in this domain for its interpretability and better robustness. To show that \toolname\ can actually help the AV security practitioners in real-life, we analyzed a vast number of potentially suspicious code blocks found from our consensus sequences, and upon analysis, we evaluated their maliciousness depending on the context. Some of them have been shared in Section \ref{sec:explainability} and \ref{app_sec:explain} (after getting double-checked by a security analyst from a reputed AV company). Besides, we also generated some statistics on our consensus sequence analysis (see \ref{app_sec:seq_analysis}).

\subsection{\textbf{Limitations}}
The approach used in \toolname\ has some limitations too. 
Firstly, though with recent and efficient multiple sequence alignment tools like SibeliaZ ~\citep{sibeliaz} we can align malware files very fast, the time can grow with the number of samples in a family. 
Secondly, any static analyzer is fundamentally vulnerable since it cannot execute and analyze the behavior of the malware. Though \toolname\ provides robustness against some specific attacks, such as gradient-based patch attacks, it might not be the case for all attacks (discussed in \ref{app_sec:robust_discussion}).  
Thirdly, like most other static analyzers, it cannot perform on packed and encrypted malware without unpacking or decrypting it.

%% file: files/conclusion.tex
In this paper, we presented a malware family classification tool \toolname\ adopting a recently developed multiple whole-genome alignment tool SibeliaZ. Sequence alignment based approaches have been used for malware analysis in the past, but the use of a whole-genome alignment tool makes \toolname\ scalable to long malware sequences and protects against trivial adversarial attacks. 
The method is also interpretable and can be used to derive insights on malware such as the identification of critical code blocks and code obfuscation. To our knowledge, we are the first to explore the interpretability on a raw-byte based static malware classifier. 

We have applied \toolname\  on the Kaggle Microsoft Malware Classification Challenge (Big 2015) and the Microsoft Machine Learning Security Evasion  Competition (2020)  (MLSec) datasets, and observed that it outperforms state-of-the-art static classifier methods such as the MalConv, Feature-Fusion, and M-CNN method. However, outperforming other models is not the main goal of this paper, rather we wanted to explore static analysis from a different perspective to balance accuracy, interpretability, and robustness. 

We discovered malicious code-blocks from our consensus sequences that can provide insights to malware analysts. We welcome security practitioners to utilize \toolname\ for not only classifying but also analyzing the malware. \toolname\ can be used in the future to measure the variance among malware samples which might provide insights about common practices by malware authors. We believe it might open up a new paradigm in the malware analysis domain.


%% file: files/app_models.tex
\section{Models}\label{app_sec:model}

\subsection{Hyperparameters for Machine Learning Models}
We experimented with the hyper-parameters for our machine learning models. The finding for each model are given below --

\subsubsection{Logistic Regression}
We got the best results for `elasticnet’ penalty, C=0.05 (regularization factor), `saga' solver and l1\_ratio=0.5.

\subsubsection{Decision Tree}
We got the best result with `gini' criterion, `best' splitter, `none' max\_depth keeping the min\_samples\_split=5.

\subsubsection{Support Vector Machine (SVM)}
We found that the SVM classifier performed best for C=0.005 (regularization factor) and `linear' kernel.
\\
\subsection{Hyperparameters for Deep Learning Models}

\subsubsection{MalConv Model}
For the MalConv model, we followed the similar implementation of the original paper \citep{malconv}. We used SGD with Nesterov momentum with $0.01$ learning rate and $0.9$ momentum. We set the max input length to $2MB$ and $8$ samples per batch. We trained the model for 20 epochs.

\subsubsection{M-CNN model}
For the M-CNN model, we used the same backbone and hyperparameters mentioned in the original paper \citep{kalash2018malware}. We used VGG-16 as the backbone setting the learning rate to $0.001$ with $0.9$ momentum. We trained the model for 20 epochs.

\subsubsection{Feature-Fusion Model}
For this model, we used the implementation that was made available on github by \cite{ahmadi2016novel} in their paper. We used `random forest' as the classifier.

\subsection{Deep Learning Model Architectures}
\begin{figure*}[h!] 
    \centering
  \subfloat[\label{1a}]{%
       \includegraphics[width=0.7\linewidth]{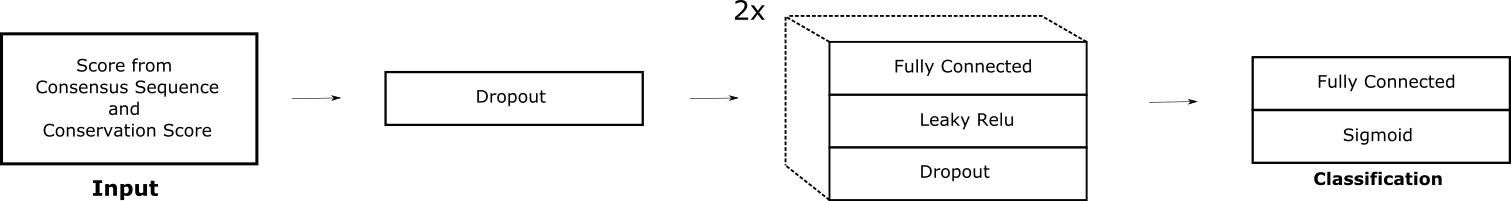}}
    \vspace{0.1cm}
 \\
  \subfloat[\label{1b}]{%
        \includegraphics[width=0.7\linewidth]{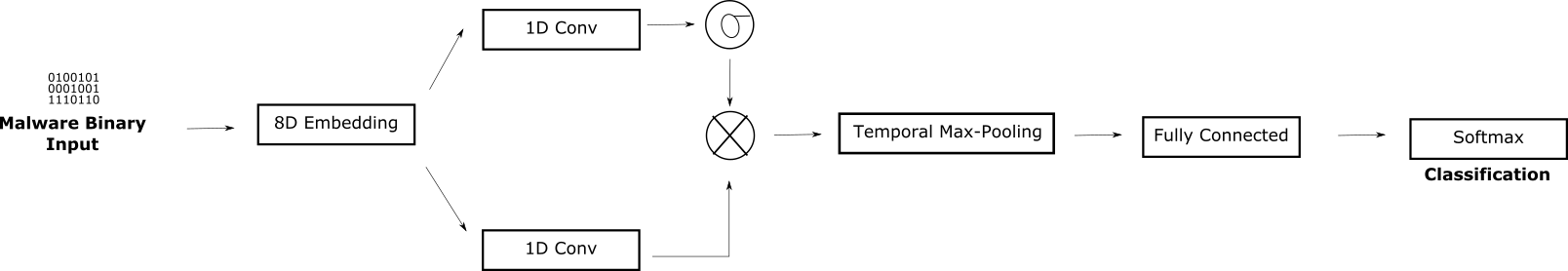}}
        \vspace{0.1cm}
 \\
  \subfloat[\label{1c}]{%
        \includegraphics[width=0.7\linewidth]{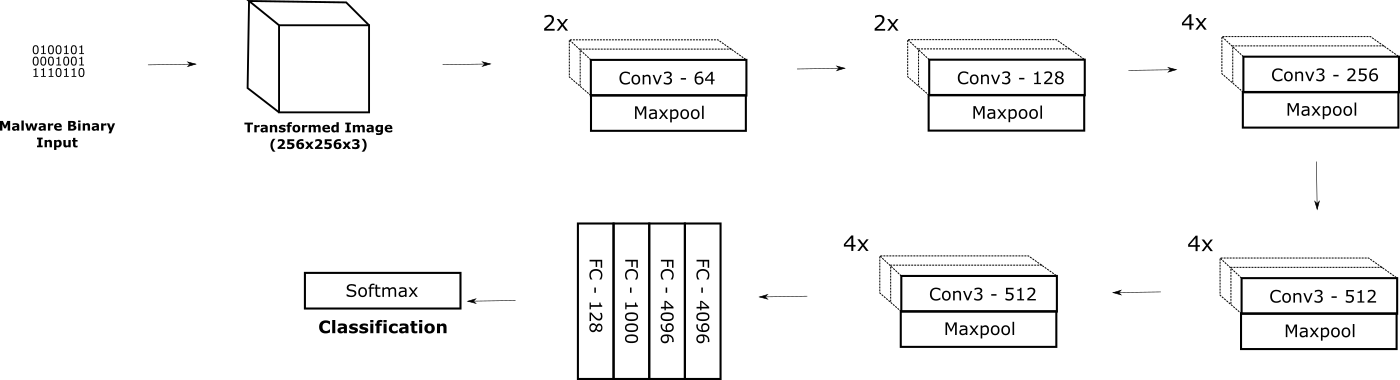}}
    \hfill
\\
  \caption{Architectures of (a) \toolname\ (Deep Learning) model, (b) MalConv model~\cite{malconv}, and (c) M-CNN model \cite{kalash2018malware}. }
  \label{fig: architectures} 
\end{figure*}

%% file: files/app_results.tex
\section{Result Tables}\label{app_sec:result}

\begin{table*}[h]
    \centering
    \resizebox{0.5\textwidth}{!}{
    \begin{tabular}{c |c c c } 
        \hline 
        \multirow{2}{*}{Family Name} & \multicolumn{3}{c}{Train Accuracy}  \\
        {} & Logistic regression & Decision tree & SVM \\
        \hline
        Ramnit & 99.91 & 99.91 & 99.91 \\ 
        Kelihos\textunderscore ver3 & 99.83 & 99.94 & 99.81 \\
        Vundo & 100 & 100 & 100  \\
        Simda & 100 & 100 & 97.92 \\
        Tracur & 100 & 100 & 99.5 \\
        Kelihos \textunderscore ver1 & 100 & 100 & 99.5 \\
        Obfuscator.ACY & 100 & 100 & 100  \\
        Gatak & 99.17 & 99.17 & 99.17  \\
        \hline
        Overall & 99.82 & 99.86 & 99.74 \\
        [0.5ex]
        \hline
    \end{tabular}
    }
    \caption{Training Accuracy on Kaggle Microsoft Malware Dataset  for Different Machine Learning Models}        
    \label{table:machine learning models result train}
\end{table*}

\begin{table*}[h!]
    \centering
    \resizebox{0.8\textwidth}{!}{
    \begin{tabular}{c |c c| c c c } 
        \hline 
        \multirow{2}{*}{Family Name} & \toolname & \toolname & \multirow{2}{*}{Feature-Fusion} & \multirow{2}{*}{MalConv} & \multirow{2}{*}{M-CNN} \\ 
        & (Logistic Regression) & (Deep Learning) & & & \\ [0.5ex] 
        \hline
        Ramnit & 99.91 & 99.58 & 100 & 98.39 & 97.64 \\ 
        Kelihos\textunderscore ver3 & 99.83 & 99.86 & 100 & 99.89 & 99.91 \\
        Vundo & 100 & 98.26 & 100 & 99.4 & 99.26 \\
        Simda & 100 & 94.44 & 100 & 100 & 100 \\
        Tracur & 100 & 98.18 & 100 & 98.74 & 98.43 \\
        Kelihos \textunderscore ver1 & 100 & 98.92 & 100 & 98.54 & 99.52 \\
        Obfuscator.ACY & 100 & 98.66 & 100 & 97.33 & 99.70 \\
        Gatak & 99.17 & 99.2 & 100 & 99.15 & 92.69 \\ 
        \hline
        Overall & 99.82 & 99.24 & 100 & 98.96 & 98.4 \\
        [0.5ex]
        \hline
    \end{tabular}
    }
    \caption{Training Accuracy of different models on Kaggle Microsoft Malware Classification Challenge Dataset}        
    \label{table:microsoft classification train accuracy for different models}
\end{table*}

\begin{table*}[h!]
    \centering
    \resizebox{\textwidth}{!}{
    \begin{tabular}{c | c  c  c | c  c  c | c  c  c} 
        \hline 
        \multirow{2}{*}{Family Name} & \multicolumn{3}{c}{\toolname } & \multicolumn{3}{|c}{\multirow{2}{*}{MalConv}} & \multicolumn{3}{|c}{\multirow{2}{*}{M-CNN}} \\
        & \multicolumn{3}{c}{(Logistic Regression)} & \multicolumn{3}{|c}{} & \multicolumn{3}{|c}{} \\
        \hline
        & Recall & Specificity & Precision & Recall & Specificity & Precision & Recall & Specificity & Precision \\
        \hline
        Ramnit & \textbf{99.65} & \textbf{99.63} & \textbf{99.65} & 96.75 & 94.81 & 94.9 & 90.26 & 86.36 & 86.88 \\ 
        Kelihos\textunderscore ver3 & 98.98 & 99.32 & 99.32 & \textbf{100} & \textbf{100} & \textbf{100} & 100 & 99.32 & 99.32 \\
        Vundo & 91.67 & \textbf{100} & \textbf{100} & 96.15 & 93.06 & 95.24 & \textbf{98.55} & 96.36 & 94.44 \\
        Simda & \textbf{100} & 71.43 & 75 & 52.94 & 0 & \textbf{100} & 54.54 & \textbf{80} & 85.71 \\
        Tracur & 89.29 & \textbf{100} & \textbf{100} & 91.62 & 98 & 98.79 & \textbf{94.49} & 94.57 & 91.96 \\
        Kelihos \textunderscore ver1 & \textbf{98.04} & 95.24 & 96.15 & 96.29 & 95.83 & 96.29 & 90.79 & \textbf{98.65} & \textbf{98.57} \\
        Obfuscator.ACY & 94.53 & \textbf{97.52} & \textbf{97.58} & \textbf{97.5} & 91.15 & 92.12 & 89.03 & 94.47 & 94.19 \\
        Gatak & 95.05 & 97.03 & 96.97 & \textbf{97.61} & \textbf{100} & \textbf{100} & 84.08 & 93.85 & 93.89 \\
        \hline
        Overall & 96.66 & \textbf{98.60} & \textbf{98.67} & \textbf{97.37} & 96.48 & 97.36 & 93.44 & 94.85 & 94.59\\
        [0.5ex]
        \hline
    \end{tabular}
    }
    \caption{Performance of different models on Test-set of Kaggle Microsoft Malware Classification Challenge Dataset for other evaluation metrics}        
    \label{table:microsoft performane test-set (all)}
\end{table*}

\begin{table}[h]
    \centering
    \resizebox{0.5\textwidth}{!}{
    \begin{tabular}{c c c c}
        \hline 
        \multirow{2}{*}{Models} & Train & Validation & Test \\ 
         & Accuracy & Accuracy & Accuracy  \\
        [0.5ex] 
        \hline
        \toolname \ (Deep Learning) & 97.53 & \textbf{81.67} & \textbf{79.58} \\
        [0.5ex]
        \hline
        Feature-Fusion & 96.83 & 58.59 & 62.50 \\
        MalConv & 95.39 & 81.22 & 64.45 \\
        M-CNN & \textbf{98.4} & 74.29 & 73.83 \\ [1ex]
        \hline
    \end{tabular}
    }
    \caption{Performance of Deep Learning Models on MLSEC Dataset for train-validation-test split}
    \label{table:MLSEC performance train-val-test}
\end{table}

%% file: files/app_explainability.tex
\section{Explainability}\label{app_sec:explain}

\subsection{Vundo Trojan}

\begin{figure*}[h]
\centering
\subfloat[\label{fig:5a}Code in .text segment (MD5 hash = jEZkrOP7GayNmiXdget5)]{\includegraphics[width = 0.75\textwidth]{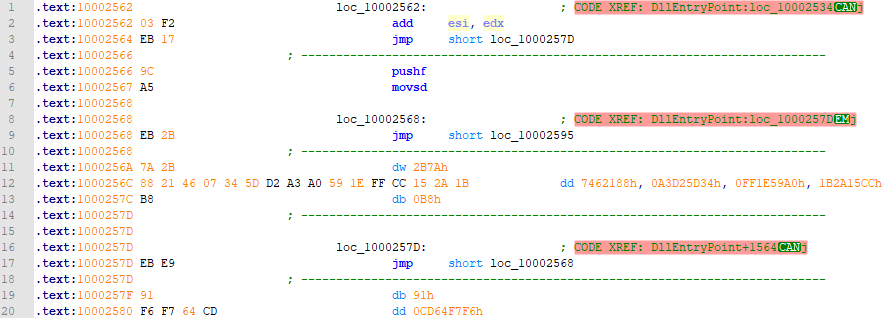}} \\
\vspace{.5cm}
\subfloat[\label{fig:5b}Code in .text segment (MD5 hash = kRUx3TuoJSgp0sqDzNGX)]{\includegraphics[width =0.7\textwidth]{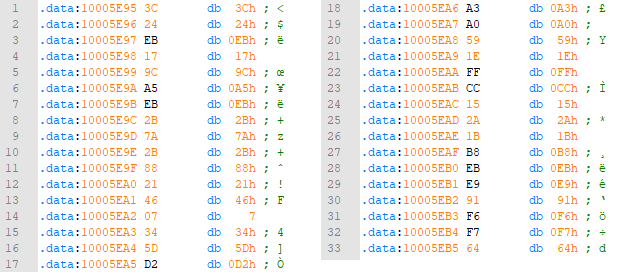}}
\caption{Code obfuscation in Vundo Trojan. \ \ (a) and (b) have the same code in \texttt{.text} and \texttt{.data} section, respectively}
\label{fig:vundo obfuscation 5}
\end{figure*}

\begin{figure*}[h]
\centering
\subfloat[\label{fig:4a}Code in .text segment (MD5 hash = g7vEfrR3s49CH8AVUeSu)]{\includegraphics[width = 0.7\textwidth]{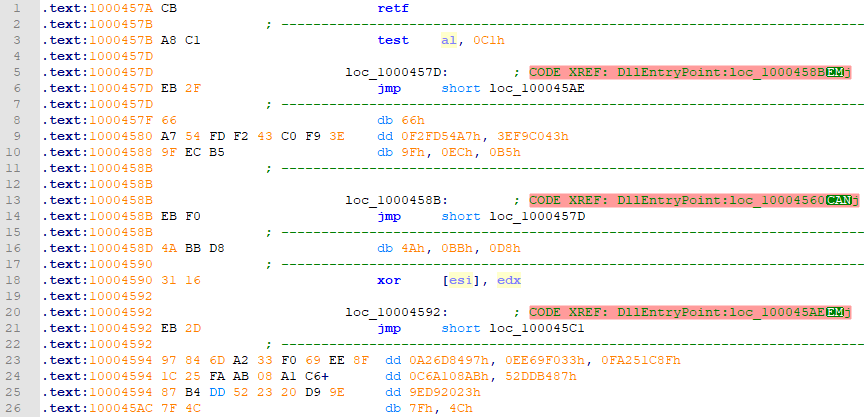}} \\
\vspace{.5cm}
\subfloat[\label{fig:4b}Code in .text segment (MD5 hash = Gpd5s3Y7HKwXf4NaOPZk)]{\includegraphics[width =0.7\textwidth]{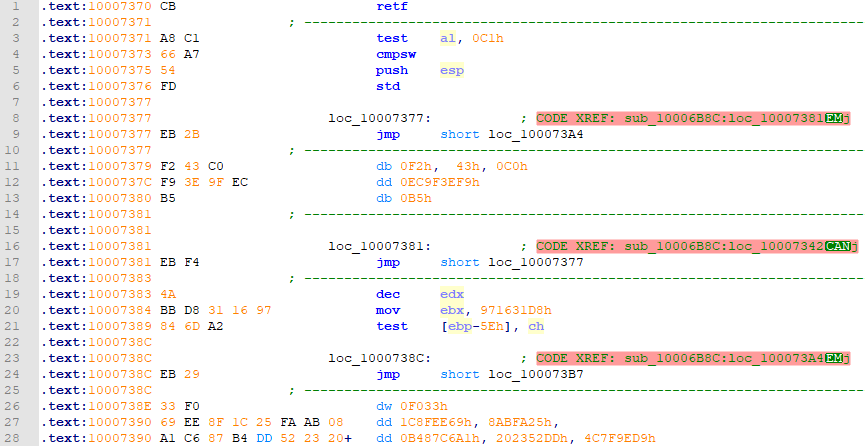}}
\vspace{.5cm}
\subfloat[\label{fig:4c}Obfuscated Code in .data segment (MD5 hash = 6WbENDkcC750euPGqApQ)]{\includegraphics[width =0.7\textwidth]{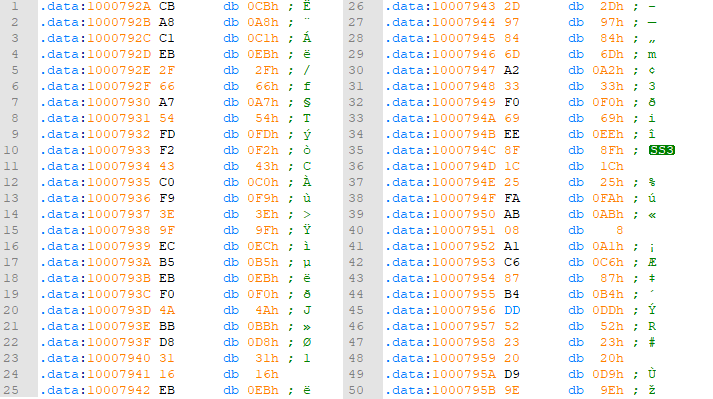}}
\caption{Code obfuscation in Vundo Trojan. \ \ (a) and (b) have the same code in \texttt{.text} section  whereas (c) obfuscated that in its \texttt{.data} section}
\label{fig:malware obfuscation 4}
\end{figure*}

We found multiple consensus sequences, i.e. signatures of \toolname, for Vundo Trojan family where code obfuscation was being used by malware authors. Here, we show two such examples in Figures \ref{fig:vundo obfuscation 5} and \ref{fig:malware obfuscation 4}. In both of them, we found the same binary sequence being used in \texttt{text} and \texttt{data} section of PE files. We suspect that -- one malware sample is loading another by keeping the code in its \texttt{data} section.

\clearpage

\subsection{Simda Backdoor}

\begin{figure*}[h]
    \centering
    \includegraphics[width=0.7\textwidth]{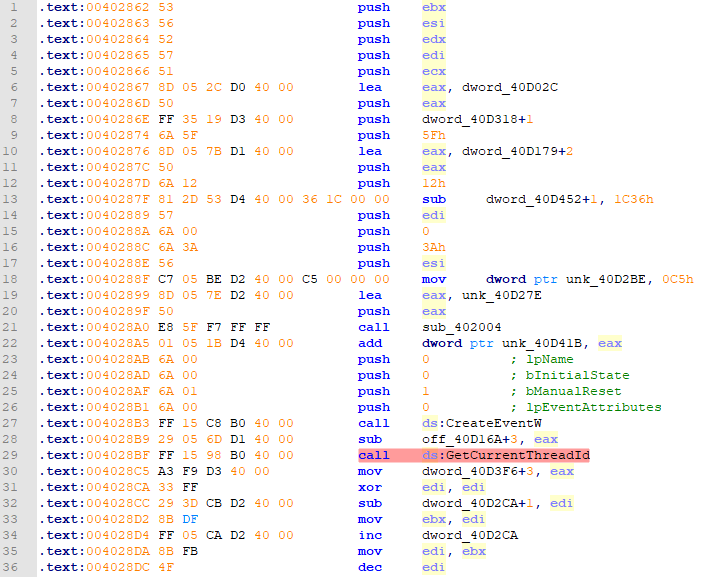}
    \caption{\texttt{GetCurrentThreadId} function call by Simda Backdoor matched with consensus sequence}
    \label{fig:simda_get_thread}
\end{figure*}

We also found that most of the malware from this family is getting matched with consensus sequence where they are calling the \texttt{\textbf{GetCurrentThreadId}} function which is one of the most important prerequisites for a backdoor (see Figure \ref{fig:simda_get_thread}), and thus, we found this function in the report by various sandboxes for previously found backdoors, e.g., JoeSandbox~\cite{joe_backdoor}, Falcon Sandbox~\cite{falcon_bakcdoor}, etc.

\clearpage

\subsection{Kelihos\_ver1 Backdoor}

\begin{figure*}[h]
\centering
\subfloat[\label{fig:kelihos1_1} First checks for admin privilege, then calls suspicious functions from IEAdvpack.dll and deletes file (0YJRabHQ1reDxN3yjUCd)]{\includegraphics[width = 0.9\textwidth]{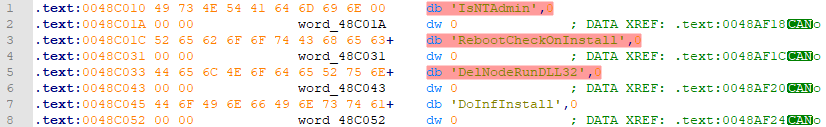}} \\
\vspace{.5cm}
\subfloat[\label{fig:kelihos1_2}First checks for admin privilege, then calls suspicious functions from IEAdvpack.dll (7cx4QdBWg9wMa2pvOLuo)]{\includegraphics[width =0.9\textwidth]{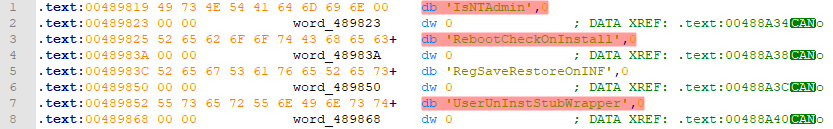}}
\vspace{.5cm}
\subfloat[\label{fig:kelihos1_3} First calls suspicious functions from IEAdvpack.dll, then checks for admin privilege (d5ki60fcbqeH3SvZzpoB)]{\includegraphics[width =0.9\textwidth]{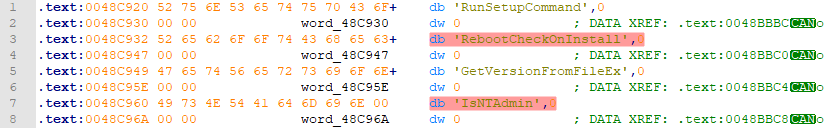}}

\caption{Suspicious Windows function calls by Kelihos (version 1) got detected by \toolname}
\label{fig:kelihos1_api_calls}
\end{figure*}

Figure \ref{fig:kelihos1_api_calls} is one of the examples where our consensus sequence got matched with byte sequence that might be related to a malicious function. But here an important observation is -- the order of the suspicious calls is different in different samples. All of them are checking for admin privileges and reboot requirement. While the malware in Figures \ref{fig:kelihos1_1} and \ref{fig:kelihos1_2} declares \texttt{IsNTAdmin} before \texttt{RebootCheckInstall}, malware in Figure \ref{fig:kelihos1_3} does not. But still, they get detected by \toolname.

\clearpage

%% file: files/app_robustness.tex
\section{Robustness}\label{app_sec:robust}

\subsection{Plausible Techniques to Evade \toolname} \label{app_sec:robust_discussion}



To evade our model, some possible attacks  and the robustness of our model against them are discussed below:
\begin{itemize}
    \item \textit{Including pieces of other malware to confuse family detection:} The inclusion of pieces from other malware will be detected but code-pieces of the original family will also be detected at the same time, and consequently, the score for the original family should be higher since it contains more code-pieces than the other family. However, if an attacker can include enough code-segments from other families that can outnumber the score for the real family, \toolname\ might predict a wrong family. One defense against this attack can be to incorporate monotonic classifier ~\cite{incer2018adversarially}, or non-negative network~\cite{fleshman2018non} with \toolname.
    \item \textit{Code Randomization: } Attackers can try to replace instructions with semantically similar ones to evade \toolname ~\cite{song2020mab}. However, the attacker will have to change a lot of instructions so that it does not get aligned with enough consensus sequences which is not a trivial task, not impossible at the same time. To defend against such attacks, we can generate artificial adversarial malware using code randomization and train \toolname\ using them to capture the signature for such code randomizations.
    \item \textit{Register Reassignment: } An attacker can add multiple instructions that reassign registers and create an adversarial malware that has different sequences than our signature~\cite{lucas2021malware}. One way to defend against this attack with \toolname\ can be -- changes in the alignment algorithm (or apply heuristic) so that it becomes flexible against register names and instructions.
    \item \textit{Instruction Reordering: } Attackers can try changing the order of instructions without harming the functionality of malware. \toolname\ is robust to such attack to some extent, and we discovered that in one of our consensus sequences (see Figure \ref{fig:kelihos1_api_calls}). However, if the order is jumbled to an extreme, possibly through the use of jump instructions, it might get evaded.\\ \\
    
\end{itemize} 

\subsection{Direct Attack on \toolname} Besides the above mentioned approaches, an attacker can try generating adversarial malware specifically for \toolname\ in a white-box setting. Since  \toolname\ is hard to invert and is non-differentiable, the attacker cannot compute gradient or apply reverse engineering unlike other end-to-end deep learning models. They will have to replace the feature mapping function, or especially the MSA tool with the best approximated model ~\cite{petti2021end} to create an adversarial sample. But replacing an MSA with approximated DNN is another challenging task. Moreover, when the attacker has to preserve the semantics, creating such an approximated model becomes even harder.

%% file: files/app_seq_analysis.tex
\section{Consensus Sequence Analysis} \label{app_sec:seq_analysis}

We analyzed the conserved sequences to estimate the number of informative code blocks for each malware family.
Table \ref{table:consensus seq count} (in Appendix) represents the number of consensus sequences, i.e., conserved blocks and the number of informative code blocks for each family in the Kaggle Microsoft dataset. We term a code block `informative' if the value of the corresponding weight in our regression model is positive. The rationale behind this is that - the higher the weight of the consensus sequence, the higher impact it has on the classification. 

We can observe from the table that, the number of consensus sequences found by \toolname\ is proportional to the number of samples in a malware family. However, the percentage of the number of informative consensus sequences varies across malware families. The percentage values indicate that  large fractions of blocks identified through the sequence alignment process are useful for the subsequent classification task.

\begin{table*}[ht!]
    \centering
    \resizebox{0.7\textwidth}{!}
    {
    \begin{tabular}{c c c c} 
        \hline 
        \multirow{2}{*}{Family Name} & Number of & Number of Informative & Percentage of Informative\\ 
        & Consensus Sequence &  Consensus Sequence & Consensus Sequence\\ [0.5ex] 
        \hline
        Ramnit & 962129 & 177905 & 18.49\% \\ 
        Kelihos\textunderscore ver3 & 331432 & 105223 & 31.75\% \\
        Vundo & 27210 & 17601 & 64.69\%\\
        Simda & 3098 & 1194 & 38.54\% \\
        Tracur & 37967 & 20548 & 54.12\% \\
        Kelihos \textunderscore ver1 & 27459 & 13467 & 49.04\% \\
        Obfuscator.ACY & 94172 & 42863 & 45.51\% \\
        Gatak & 275984 & 96554 & 34.99\% \\
        [0.5ex]
        \hline
    \end{tabular}
    }
    \caption{Number of Consensus Sequence in Malware Families}        
    \label{table:consensus seq count}
\end{table*}